\documentclass[sigconf]{acmart}

\usepackage{multirow}
\usepackage{subfigure}
\usepackage{amsmath}
\usepackage{amsthm}
\usepackage{amsfonts} 
\usepackage[ruled,linesnumbered]{algorithm2e}
\usepackage{booktabs}
\usepackage{graphicx}
\usepackage{bm}
\usepackage{enumitem}
\usepackage{xcolor}
\usepackage{makecell}
\usepackage{bbding}
\usepackage{balance}

\newcommand{\bW}{\mathbf{W}}
\newcommand{\ms}[2]{$#1_{(#2)}$}
\newcommand{\bms}[2]{$\mathbf{#1}_{(#2)}$}
\newcommand{\sbms}[2]{$\underline{#1}_{(#2)}$}
\newcommand{\TheName}[0]{\textbf{PERSCEN}}

\AtBeginDocument{%
  }

\copyrightyear{2025}
\acmYear{2025}
\setcopyright{acmlicensed}
\acmConference[KDD '25] {Proceedings of the 31st ACM SIGKDD Conference on Knowledge Discovery and Data Mining V.2}{August 3--7, 2025}{Toronto, ON, Canada.}
\acmBooktitle{Proceedings of the 31st ACM SIGKDD Conference on Knowledge Discovery and Data Mining V.2 (KDD '25), August 3--7, 2025, Toronto, ON, Canada}
\acmISBN{979-8-4007-1454-2/25/08}
\acmDOI{10.1145/3711896.3737079}
\begin{document}

\title{PERSCEN: Learning Personalized Interaction Pattern and Scenario Preference for Multi-Scenario Matching}

\author{Haotong Du}
\affiliation{%
	\department{School of Computer Science}
	\institution{Northwestern Polytechnical University}
	\city{Xi'an}
	\state{Shaanxi}
	\country{China}
}
\email{duhaotong@mail.nwpu.edu.cn}

\author{Yaqing Wang}
\authornote{Corresponding authors.}
\affiliation{%
	\institution{Beijing Institute of Mathematical
		Sciences and Applications}
	\city{Beijing}
	\country{China}}
\email{wangyaqing@bimsa.cn}

\author{Fei Xiong}
\affiliation{%
	\institution{Meituan}
	\city{Beijing}
	\country{China}}
\email{xiongfei07@meituan.com}

\author{Lei Shao}
\affiliation{%
	\institution{Meituan}
	\city{Beijing}
	\country{China}}
\email{shaolei05@meituan.com}

\author{Ming Liu}
\affiliation{%
	\institution{Meituan}
	\city{Beijing}
	\country{China}}
\email{liuming66@meituan.com}

\author{Hao Gu}
\affiliation{%
	\institution{Meituan}
	\city{Beijing}
	\country{China}}
\email{guhao02@meituan.com}

\author{Quanming Yao}
\affiliation{%
	\department{Department of Electronic Engineering}
	\institution{Tsinghua University}
	\city{Beijing}
	\country{China}}
\email{qyaoaa@tsinghua.edu.cn}

\author{Zhen Wang}
\authornotemark[1]
\affiliation{%
	\department{School of Cybersecurity}
	\institution{Northwestern Polytechnical University}
	\city{Xi'an}
	\state{Shaanxi}
	\country{China}
}
\email{w-zhen@nwpu.edu.cn}
\renewcommand{\shortauthors}{Haotong Du et al.}
\begin{abstract}
With the expansion of business scales and scopes on online platforms, 
multi-scenario matching has become a mainstream solution to reduce maintenance costs and alleviate data sparsity. 
The key to effective multi-scenario recommendation lies in capturing both user preferences shared across all scenarios and scenario-aware preferences specific to each scenario.
However, 
existing methods often overlook user-specific modeling, 
limiting the generation of personalized user representations. 
To address this, 
we propose PERSCEN, 
an innovative approach that incorporates user-specific modeling into multi-scenario matching.
PERSCEN constructs a user-specific feature graph based on user characteristics
and employs a lightweight graph neural network to capture higher-order interaction patterns, 
enabling personalized extraction of preferences shared across scenarios. 
Additionally, 
we leverage vector quantization techniques to distil scenario-aware preferences from users' behavior sequence within individual scenarios, 
facilitating user-specific and scenario-aware preference modeling. 
To enhance efficient and flexible information transfer, 
we introduce a progressive scenario-aware gated linear unit that allows fine-grained, 
low-latency fusion. 
Extensive experiments demonstrate that PERSCEN outperforms existing methods. 
Further efficiency analysis confirms that PERSCEN effectively balances performance with computational cost,
ensuring its practicality for real-world industrial systems.
\end{abstract}


\maketitle

\section{Introduction}

As online platforms diversify their services and expand content ecosystems, 
multi-scenario matching becomes a key focus in recommendation systems, 
aiming to efficiently retrieve high-quality items from large-scale candidate pools across diverse scenarios.

\begin{figure}[ht]
	\centering
	\includegraphics[width=0.49\textwidth]{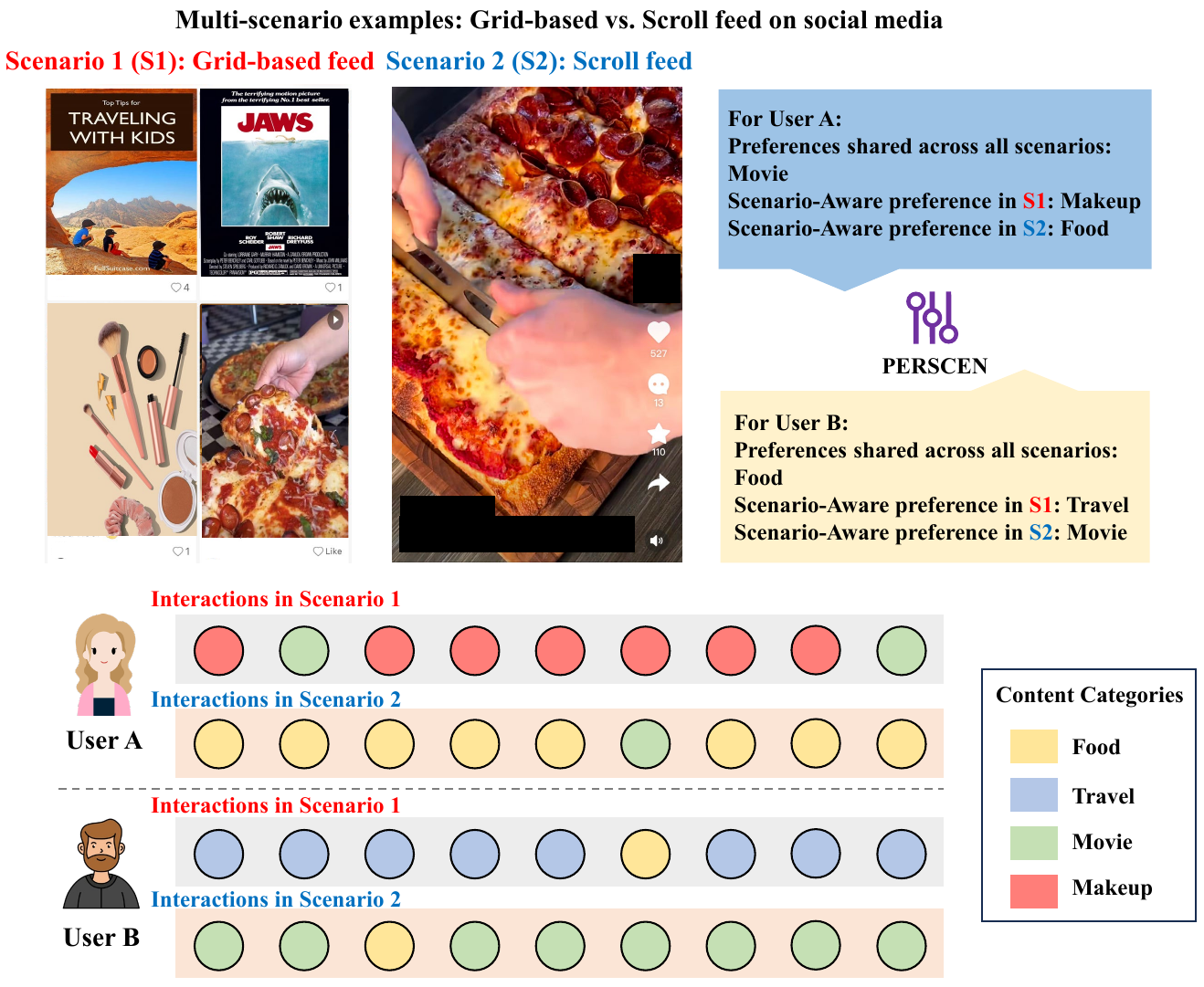}
	\caption{
		For different users, preferences shared across all scenarios and scenario-aware preferences are unique. User A’s shared preference is movies, while User B’s is food. Scenario-aware preferences shift with feed layout: in grid-based feeds, User A favors makeup and User B prefers travel; in scroll feeds, User A shifts to food and User B to movies. Therefore, user-specific modeling is crucial for effective multi-scenario recommendation.
	}
	\label{fig:motivation}
	\vspace{-5px}
\end{figure}

In multi-scenario recommendation systems, 
users often exhibit consistent preferences across different content delivery scenarios—such as homepage feeds, banner sections, or scrollable content streams. 
These preferred scenarios may share common behavioral patterns driven by user interests. 
However, each scenario also presents distinct contextual characteristics (e.g., layout, content type, interaction modality) that significantly influence user behavior. 
This interplay of preferences shared across all scenarios and contextual differences is crucial for building accurate and personalized recommendation models that can generalize across diverse business scenarios.
Existing multi-scenario recommendation methods focus on modeling this interplay. 
By employing various modulation techniques,
such as attention mechanisms~\cite{xie2020internal}, 
dynamic weighting~\cite{jiang2022adaptive}, 
and gating mechanism~\cite{jiang2022adaptive,zhao2023m5},
these approches adaptively transfer shared preferences across different scenarios, 
achieving effective multi-scenario recommendations. 
However, 
these approaches often neglect fine-grained, 
user-specific modeling, 
instead relying on scenario context features that fail to capture 
the nuanced behavioral differences of individual users across scenarios.
This limitation can lead to less accurate recommendations and suboptimal user experiences, 
especially considering that user-specific modeling aligns more closely with real-world user behavior, 
where users engage with different scenarios in distinct ways, as shown in Figure~\ref{fig:motivation}.

Nevertheless, capturing user-specific preferences requires fine-grained and often resource-intensive modeling. 
In contrast, the matching stage must efficiently retrieve candidates from millions of items under strict latency and resource constraints, rendering the direct use of heavy-weight models impractical. 
As a result, enabling user-specific modeling in multi-scenario matching remains a significant technical challenge.

To bridge this gap, 
we propose PERSCEN, 
an efficient multi-scenario matching model centered on user-specific modeling. 
Specifically, 
PERSCEN constructs a user-specific feature graph and leverages a lightweight graph neural network (GNN) to capture user-specific interaction patterns.
To effectively model contextual differences, 
we introduce a vector quantization-based extraction mechanism that distils user-specific and scenario-aware preference from scenario-aware behavior sequences. 
Finally, 
to efficiently fuse shared preferences and contextual differences, 
we design a progressive scenario-aware gated linear unit (GLU) that enables adaptive and efficient information integration.
Note the PERSCEN caters to individual user needs across diverse scenarios while maintaining high retrieval efficiency.

Our contributions can be summarized as follows: 
\begin{itemize}[leftmargin=*] 
	\item To the best of our knowledge, PERSCEN is the first to incorporate user-specific modeling into multi-scenario matching, enabling fine-grained extraction of multi-scenario information tailored to individual users.
	\item Technically, PERSCEN captures preferences shared across all scenarios via a lightweight GNN on a user-specific feature graph, models user-specific and scenario-aware preferences through vector quantization of scenario-aware behaviors, and integrates both via a progressive scenario-aware GLU, achieving personalized multi-scenario matching with low computational cost. 
	\item We perform extensive experiments on multiple benchmark datasets, demonstrating that PERSCEN outperforms existing approaches. Furthermore, our efficiency analysis confirms that PERSCEN achieves a compelling balance between recommendation performance and computational cost, underscoring its practicality for deployment in real-world industrial systems.
\end{itemize}

\begin{figure*}[h]
	\centering
	\includegraphics[width=0.98\textwidth]{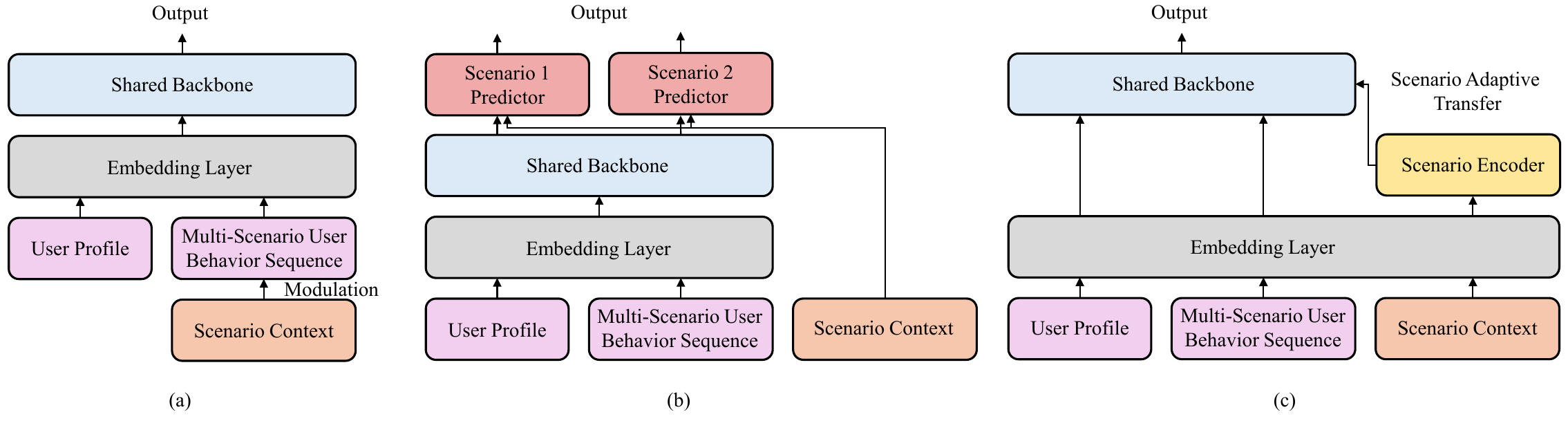}
	\caption{Overview of existing strategies for multi-scenario matching. These methods take user profile, scenario context, and multi-scenario user behavior sequences as input. A shared embedding layer and backbone are typically used to produce the final matching prediction. Subfigures (a), (b), and (c) illustrate representative designs that integrate different modules to model scenario distinctions. For detailed descriptions of each strategy, please refer to the corresponding discussion in the main text.
	}
	\label{fig:existing_works}
\end{figure*}
\section{Related Works}

Recommendation systems~\cite{bian2023feynman,wu2023coldnas,liu2024knowledge} typically operate in two primary stages: matching and ranking. 
In this paper, we focus on the multi-scenario matching problem~\cite{xie2020internal, jiang2022adaptive, zhao2023m5, zhang2022scenario}, 
which aims to efficiently identify the most relevant items for users across diverse scenarios from a vast candidate pool.

For the matching stage,
two-tower models have become a widely adopted architecture~\cite{huang2020embedding}, 
particularly for retrieving a preliminary list of candidates from a large item corpus based on a user's historical interactions.
These models consist of two parallel deep neural networks
that independently learn embeddings for users and items. 
The resulting embeddings are then compared using efficient similarity calculations within an Approximate Nearest Neighbor (ANN) retrieval system, enabling fast candidate retrieval~\cite{johnson2019billion}.
This architecture is highly valued for its high throughput and low latency, 
making it particularly suitable for industrial applications.
Building on the foundation of two-tower models, 
DSSM~\cite{huang2013learning} maps queries and documents into 
a shared semantic space, trained by maximizing the cosine similarity between their semantic vectors to facilitate effective retrieval.
MIND~\cite{li2019multi} introduces dynamic routing from capsule networks to model the multiple facets of a user's interests, 
extracting diverse representations from their raw behavior sequences.
DAT~\cite{yu2021dual} enhances the interaction between 
the user and item towers by incorporating dual enhancement vectors, 
enabling implicit information exchange.

Extending these two-tower frameworks,
existing multi-scenario matching methods typically concentrate on modeling how preferences shared across all scenarios interact with scenario-specific differences, 
as illustrated in Figure~\ref{fig:existing_works}. 
ICAN~\cite{xie2020internal} (Figure~\ref{fig:existing_works}(a)) introduces a scenario-aware internal and contextualized attention mechanism to modulate the multi-scenario user behavior sequence
before feeding it into a shared backbone for prediction.
ADIN~\cite{jiang2022adaptive} (Figure~\ref{fig:existing_works}(b)) 
utilizes a shared backbone but incorporates separate scenario predictors to capture both shared preferences and the unique characteristics of different scenarios.
SASS~\cite{zhang2022scenario} (Figure~\ref{fig:existing_works}(c)) 
augments the model with a scenario encoder to capture scenario-specific differences and employs a multi-layer scenario-adaptive transfer module to guide the shared backbone, 
demonstrating improved performance in certain contexts
M5~\cite{zhao2023m5} similarly leverages scenario context to modulate behavior sequences using a gating mechanism, 
and uses a split mixture-of-experts to capture shared preferences across different scenarios.
While these methods are effective at capturing general scenario patterns, they often overlook the critical aspect of user-specific modeling. Many of them employ a uniform feature interaction pattern across all users, limiting their ability to model nuanced, individualized preferences shared across all scenarios. Furthermore, they frequently rely on generic behavior sequences rather than explicitly modeling user-specific, scenario-aware preferences, a crucial element for enhancing the quality of multi-scenario matching.

The multi-scenario ranking problem is related but fundamentally different from the matching task~\cite{shen2021sar, zhou2023hinet, chang2023pepnet, tian2023multi, li2023hamur, min2023scenario, zhang2024scenario, zhu2024m, liu2024multifs, gao2024hierrec, wang2024llm4msr}. Ranking models focus on fine-grained prediction among a small set of candidates by modeling user-item interactions in detail at the input layer. In contrast, matching aims to efficiently and economically retrieve the most relevant items from a large candidate pool, making these ranking methods unsuitable for deployment in the matching stage. 

\begin{figure*}[hbtp]
	\centering
	\includegraphics[width=0.99\textwidth]{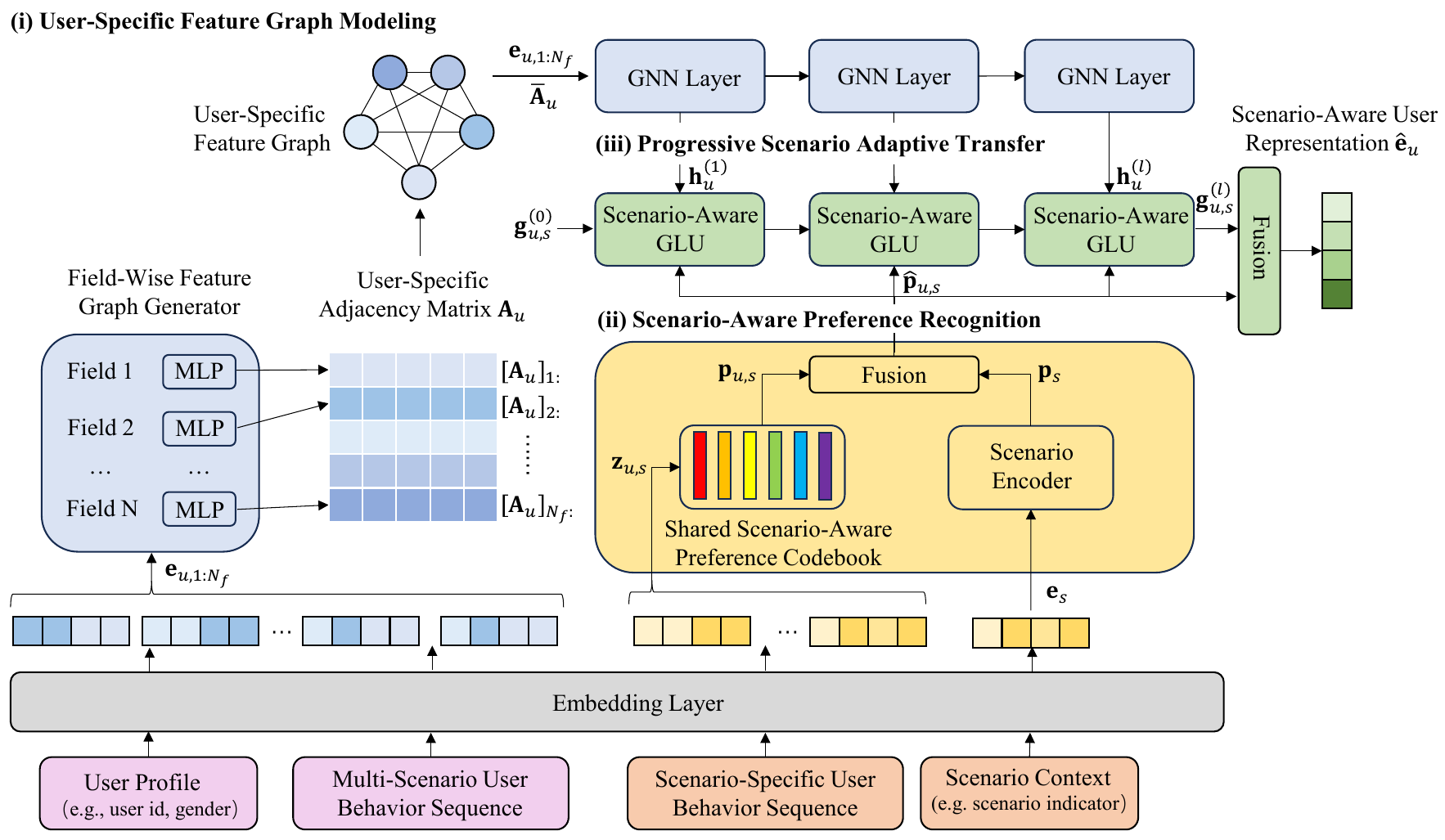}
	\caption{Framework of PERSCEN. The framework consists of three key components: 
		(i) User-Specific Feature Graph Modeling, where user-specific adjacency matrices are produced by field-wise feature graph generator to record user-specific interaction patterns; 
		(ii) Scenario-Aware Preference Recognition, where scenario-specific user behavior sequence and scenario context feature are integrated to model user-specific and scenario-aware preference jointly; and
		(iii) Progressive Scenario Adaptive Transfer which facilitates flexible information transfer across different scenarios. 
	}
	\label{fig:illus}
\end{figure*}

\section{Problem Definition}
In this paper, we consider 
multi-scenario matching problem, 
which aims to retrieval items of interest for users from a vast pool of candidate items across different scenarios.

Given a set of users $\mathcal{U}=\{u\}$, a set of items $\mathcal{V}=\{v\}$,
and a set of scenarios $\mathcal{S}=\{s\}$, 
the multi-scenario matching problem can be formulated as follows:
\begin{equation}
	\mathcal{R}_{u,s} = \underset{v\in\mathcal{V}}{\arg\text{Top}_K} g_\theta(v|u,s),
\end{equation}
where
$\mathcal{R}_{u,s}$ is the top $K$ matching results returned for user $u$ in scenario $s$, 
and $g_{\theta}$ denotes the
model parameterized by $\theta$. 

It is important to distinguish this setting from the well-studied cross-domain~\cite{ijcai2021p639} and multi-task~\cite{wang2023multi} recommendation problems.
In multi-scenario recommendation, there is no hierarchy or superiority among scenarios; the goal is to improve performance across all scenarios simultaneously. In contrast, cross-domain recommendation transfers knowledge from a source domain to enhance performance in a target domain, often dealing with challenges such as domain heterogeneity and adaptation. Meanwhile, multi-task recommendation jointly learns multiple related tasks (e.g., CTR and CVR prediction) by sharing representations, with the aim of improving generalization through task-level collaboration and regularization. 

\section{The PERSCEN Method}

In this section, we introduce PERSCEN (Figure~\ref{fig:illus}), 
an efficient framework designed to capture both 
preferences shared across all scenarios and scenario-aware preferences.
Specifically, PERSCEN includes three key components:
(i) \textbf{user-specific feature graph modeling}
(Section~\ref{sec:per_gnn})
constructs user-specific feature graphs,
and a lightweight graph neural network (GNN) is applied to these graphs to capture complex and user-specific interaction patterns; 
(ii) \textbf{scenario-aware preference recognition} (Section~\ref{sec:per_scen})
which identifies and models user-specific and scenario-aware preferences using a vector quantization-based mechanism;
and 
(iii) \textbf{progressive scenario adaptive transfer} (Section~\ref{sec:prog_sgu}) which 
enables flexible information transfer across different scenarios, optimizing performance in varying contexts. 
In the sequal, 
we primarily focus on the design of the user tower. 
Unless otherwise specified, 
the item tower follows a similar structure.
\subsection{Embedding Layer}
Consider a user $u$, 
each user feature is mapped into a low-dimensional dense vector through an embedding layer.
Specifically,
user features can be divided into sparse features~(e.g., user id, gender), 
dense features~(e.g., register days), 
and sequential features~(e.g., multi-scenario and scenario-specific user behavior sequence). 
For the $m$th feature $f_{u,m}$ of user $u$, 
its feature embedding $\mathbf{e}_{u,m}$ can be obtained as 
\begin{align}\label{eq:emb-layer}
\mathbf{e}_{u,m}= 
\begin{cases}
	\mathbf{W}_{u,m} \cdot \text{one-hot}(f_{u,m})& \text{if $f_{u,m}$ is sparse feature}\\
	\mathbf{W}_{u,m} \cdot f_{u,m}& \text{if $f_{u,m}$ is dense feature}\\
	\text{POOLING}(f_{u,m})&\text{if $f_{u,m}$ is sequence feature}
\end{cases}, 
\end{align}
where $\mathbf{W}_{u,m}$ denotes the embedding parameter of the $m$th feature field,
$\text{one-hot}(f_{u,m})$ denotes the one-hot vector of sparse feature $f_{u,m}$, 
and $\text{POOLING}(\cdot)$ denotes the sequence pooling operation, 
such as maximum pooling and average pooling.

Similarly, 
consider a scenario $s$,
each scenario feature is also mapped to low-dimensional vectors. 
For example, scenario context $f_s$ (i.e., scenario indicator), 
its embedding $\mathbf{e}_s$ is obtained as 
\begin{align}
	\label{eq:s}
	\mathbf{e}_s = \mathbf{W}_{s} \cdot \text{one-hot}(f_{s}). 
\end{align}
\subsection{User-Specific Feature Graph Modeling}\label{sec:per_gnn}
Existing approaches~\cite{jiang2022adaptive, zhang2022scenario} typically concatenate the generated feature embeddings and feed them into deep neural networks or other shared backbones to learn universal feature interaction patterns across different scenarios. 
However, 
this strategy often fails to capture the nuanced feature interaction patterns specific to individual users, thereby limiting the model’s personalization capabilities. 
In contrast, we propose generating user-specific feature graphs, 
leveraging lightweight GNNs to capture higher-order interactions between features. 
This method allows the shared backbone to maintain its personalization abilities while preserving the simplicity and efficiency inherent in the two-tower architecture.

\subsubsection{Field-Wise Feature Graph Generator}

Inspired by EmerG~\cite{wang2024warming},
we design a field-wise feature graph generator to construct a unique user-specific feature graph for each user $u$. 
In this graph, each node represents a user feature $u_m$, and the edges between nodes encode their interactions within a learned user-specific adjacency matrix. 
This approach facilitates a personalized depiction of feature interactions from each user's perspective.

Specifically, let 
$\mathbf{A}^{(1)}_{u}$ denote the user-specific adjacency matrix, 
a field-wise feature graph generator corresponding to the $m$th field
produces the $m$th row of $\mathbf{A}^{(1)}_{u}$ 
to record the connection relationship between node $m$  and the other nodes in the feature graph as:
\begin{equation}\label{eq:generator}
	[\mathbf{A}^{(1)}_{u}]_{m:} = {\text{MLP}}_{m}( [\mathbf{e}_{u,1},\dots,\mathbf{e}_{u,m},\dots,\mathbf{e}_{u,N_f},\text{one-hot}(m)]), 
\end{equation}
where ${\text{MLP}}_{m}$ denotes a multi-layer perceptron (MLP) that models the connection relationship between the $m$th feature field and other feature fields, 
$\mathbf{e}_{u,m}$ is the feature embedding obtained by \eqref{eq:emb-layer}, 
$N_f$ is the number of user profile features, 
and $[]$ represents concatenation.

The higher-order adjacency matrix can be recursively generated using matrix product,
i.e., $\mathbf{A}^{(l)}_{u}= \mathbf{A}^{(l-1)}_{u} \cdot \mathbf{A}^{(1)}_{u}$.
Furthermore, 
we refine each $\mathbf{A}^{(l)}_{u}$ using normalization, sparsification, and symmetrization to preserve important feature interactions while enhancing efficiency~\cite{wang2024warming}. 
Afterwards, we obtain
the final user-specific adjacency matrix $\bar{\mathbf{A}}^{(l)}_{u}$.
Based on the proposed field-wise feature graph generator, 
we can effectively represent user-specific feature interaction patterns for different users. 

\subsubsection{Feature Interaction on User-Specific Feature Graph}
Based on the generated user-specific feature graph, 
follow~\cite{wang2024warming},
we adopt an efficient and expressive 
GNN to capture higher-order interaction patterns on the feature graph:
\begin{align}
	\label{eq:gnn-update}
	\mathbf{h}_{u,m}^{(l)}
	=
	\mathbf{h}_{u,m}^{(l-1)}\odot
	\left(
	\sum\nolimits_{n = 1}^{N_f}{[\bar{\mathbf{A}}^{(l - 1)}_{u}]_{mn} \mathbf{W}^{(l - 1)}_g\mathbf{h}_{u,n}^{(0)}}
	\right),
\end{align}
where 
$\mathbf{h}_{u,m}^{(0)}=\mathbf{e}_{u,m}$, 
$\odot$ is the element-wise product,  
$\bar{\mathbf{A}}^{(l - 1)}_{u}$ is the final user-specific adjacency matrix, 
$[\bar{\mathbf{A}}^{(l - 1)}_{u}]_{mn}$ is the learned connection relationship between node $m$ and $n$ in the feature graph, 
and $\mathbf{W}^{(l - 1)}_g$ is a learnable parameter. 
Therefore, the $l$th layer hidden representation $\mathbf{h}^{(l)}_{u}$ of user $u$ can be denoted as:
\begin{align}
	\label{eq:user_hidden}
	\mathbf{h}^{(l)}_u = [\mathbf{h}_{u,1}^{(l)},\cdots,\mathbf{h}_{u,m}^{(l)},\dots,\mathbf{h}_{u,N_f}^{(l)}].
\end{align}

After $L$ layers of interaction modeling on the user-specific feature graph, 
the learned $\mathbf{h}^{(L)}_u$
can capture shared user preferences across all scenarios.
It's important to note that our lightweight GNN design avoids traditional message passing, 
relying instead on efficient matrix multiplications.
This approach, with its chained operations, 
significantly reduces storage costs for higher-order interactions. 
This crucial balance between efficiency and user-specific modeling is key to achieving effective multi-scenario matching.

\subsection{Scenario-Aware Preference Recognition}\label{sec:per_scen}
Previous multi-scenario recommendation works~\cite{jiang2022adaptive, zhang2022scenario, li2023hamur, gao2024hierrec} often model scenario differences using only generic scenario context feature, 
resulting in a unified preference representation for all users within a given scenario. 
However, 
individual users within the same scenario possess unique preferences that should be personalized.

To address this limitation, 
we propose modeling user-specific and scenario-aware preferences by leveraging users' scenario-specific behavior sequences. 
Inspired by recent work~\cite{yan2024trinity}, 
we apply vector quantization~\cite{liu2024learning,liu2024vector} to distil these scenario-aware preferences for individual users. 
Code vectors within a shared codebook, 
representing diverse user preferences across various scenarios,
are learned from interaction data across all scenarios. 
Crucially, because this codebook is shared, 
preferences learned in data-rich scenarios can naturally transfer to data-sparse ones. 
This elegantly overcomes the challenge faced by existing methods, 
which struggle to learn meaningful scenario-specific distinctions when data is scarce.

Specifically, for user $u$ in scenario $s$, 
we first select its scenario-specific user behavior sequence $f^{seq}_{u,s}$ 
from all their multi-scenario user behavior sequence $f^{seq}_u$. 
Then, 
we apply a sequence pooling strategy 
and encode $f^{seq}_{u,s}$ into a latent representation $\mathbf{z}_{u,s}$ by MLP:
\begin{align}
	\label{eq:scenario-sq}
	\mathbf{z}_{u,s} = \text{MLP}_{\text{proj}}(\text{POOLING}(f^{seq}_{u,s})).
\end{align}
The scenario-aware preference is recognized as 
\begin{align}
	\label{eq:vq}
	j = {\arg \min}_{i} ||\mathbf{z}_{u,s} - \mathbf{c}_i||_2^2,
\end{align}
where $\mathbf{c}_i$
is candidate code vector in the codebook,
$j$ is the index of the matched code vector $\mathbf{c}_j$.
Through vector quantization, 
we obtain the scenario-aware preference representation $\mathbf{c}_j$ that best matches the user’s historical behavior sequence in specific scenario. 
The final user scenario-aware preference is then derived through a residual connection as
\begin{align}
	\label{eq:p_dyn}
	\mathbf{p}_{u,s} = \mathbf{z}_{u,s} + \mathbf{c}_j.
\end{align}
As shown in~\eqref{eq:vq},
the vector quantization process is non-differentiable, 
we adopt the straight-through estimator~\cite{bengio2013estimating} to facilitate effective and smoother optimization of the codebook.
Therefore,
the vector quantization loss is expressed as follows:
\begin{align}
	\label{eq:loss_vq}
	\mathcal{L}_\text{VQ} = ||\text{sg}[\mathbf{z}_{u,s}] - \mathbf{c}_j||_2^2 + \beta ||\mathbf{z}_{u,s} - \text{sg}[\mathbf{c}_j]||_2^2,
\end{align}
where the first term is a codebook loss, which updates the codebook embeddings,
and the second term is a commitment loss, which promotes the hidden vector to approach the currently matched code vector. $\text{sg}[\cdot]$ denotes stop-gradient operation, and $\beta$ is a hyper-parameter that controls the trade-off between the two losses. 

For scenario context feature $f_{s}$ for scenario $s$,
we first obtain its feature embedding $\mathbf{e}_{s}$ by~\ref{eq:s}, 
and
simply use a scenario encoder to obtain the scenario representations: 
\begin{align}
	\label{eq:p_sta}
	\mathbf{p}_{s} =  \text{MLP}_{\text{scen}}(\mathbf{e}_{s}).
\end{align}
Finally, we fuse the two parts via an MLP to generate the final scenario-aware preference $\hat{\mathbf{p}}_{u,s}$:
\begin{align}
	\label{eq:personalized_scenario_pref}
	\hat{\mathbf{p}}_{u,s}= \text{MLP}_{\text{fusion}}([\mathbf{p}_{u,s}, \mathbf{p}_{s}]).
\end{align}

\subsection{Progressive Scenario Adaptive Transfer}\label{sec:prog_sgu}
Building upon the GNN-derived shared preferences and VQ-derived scenario-aware preferences, 
our goal is to fuse them adaptively.
While existing methods like GRU~\cite{cho2014learning} have been used for information transfer and fusion~\cite{jiang2022adaptive},
their complex gating mechanisms (reset and update gates) often limit flexibility and responsiveness to dynamic scenario changes, 
hindering the efficiency crucial for industrial matching models.

To overcome these limitations, 
we propose a progressive scenario-aware GLU, 
inspired by GLU~\cite{dauphin2017language} and PLE~\cite{tang2020progressive}, 
for more flexible and efficient scenario transfer. 
Unlike GRU, 
GLU utilizes simple element-wise multiplication for finer-grained control over information transfer, significantly reducing computational complexity. This inherent simplicity offers greater flexibility and adaptability, which is particularly beneficial for multi-scenario matching tasks.

The proposed module can be expressed as:
\begin{align}\notag
	\mathbf{g}^{(l)}_{u,s} = &(\bW_{r1}^{(l)}[\mathbf{h}^{(l)}_u ,\mathbf{g}^{(l-1)}_{u,s} ]+\bW_{r2}^{(l)}\hat{\mathbf{p}}_{u,s})\\	\label{eq:asgu}
	&\otimes\sigma(\bW_{r3}^{(l)}[\mathbf{h}^{(l)}_u ,\mathbf{g}^{(l-1)}_{u,s}]+\bW_{r4}^{(l)}\hat{\mathbf{p}}_{u,s}),
\end{align}
where $\mathbf{h}^{(l)}_u$ represents the user hidden representation generated by the $l$th GNN layer, and
$\mathbf{g}^{(l)}$ is the output of the $l$th scenario-aware GLU.
$\bW_{r1}^{(l)}$ and $\bW_{r2}^{(l)}$ are the project matrix and bias matrix of raw signals.
$\bW_{r3}^{(l)}$ and $\bW_{r4}^{(l)}$ are the project matrix and bias matrix of masking signals.
$\sigma(\cdot)$ is the sigmoid function, 
and $\hat{\mathbf{p}}_{u,s}$ denotes the scenario-aware preference.
Specifically,
$\mathbf{g}^{(0)}_{u,s}=\mathbf{h}^{(0)}_u$, which is the concatenation of user feature embedding as in \eqref{eq:user_hidden}.
Note that in \eqref{eq:asgu}, the sigmoid function controls the information flow, deciding which data to retain or discard. The GLU dynamically modulates the common information $\mathbf{h}^{(l)}$ from all scenarios and the distinct information $\hat{\mathbf{p}}$ based on their characteristics, allowing for more flexible fusion of scenario-specific details. 

Corresponding to GNN layers, we apply $L$-layer scenario-aware GLU, 
the final output $\mathbf{g}^{(l)}$
of the scenario-aware GLU 
is merged with the scenario-aware preference representation $\hat{\mathbf{p}}_{u,s}$ to obtain the scenario-aware user representation $\hat{\mathbf{e}}_u$:
\begin{align}
	\label{eq:final_fusion1}
	\alpha & = \sigma(\bW_o[\mathbf{g}^{(l)}_{u,s},\hat{\mathbf{p}}_{u,s}]), \\
	\label{eq:final_fusion2}
	\hat{\mathbf{e}}_u & = \alpha \cdot \mathbf{g}^{(l)}_{u,s} + (1-\alpha) \cdot \hat{\mathbf{p}}_{u,s},
\end{align}
where $\bW_o$ is a learnable parameter.
Similarly, 
we can also obtain the scenario-aware 
item representation $\hat{\mathbf{e}}_v$.

\subsection{Optimization}
The matching score for 
a user-item pair $(u,v)$ 
 is calculated as:
\begin{align}
	\label{eq:score}
	\hat{y} = \sigma(\langle \hat{\mathbf{e}}_u,\hat{\mathbf{e}}_v \rangle),  
\end{align}
where $\langle \cdot, \cdot \rangle$ represents the inner product operation.

During optimization, 
we treat the multi-scenario matching task as a binary classification problem,
and apply a random negative sampling strategy. 
Specifically,
for each positive sample, 
we randomly select negative items from the item candidate set to generate corresponding negative samples, 
which are then added to the training set. 
Therefore, the task loss can be expressed as:
\begin{align}
	\label{eq:loss_task}
	\mathcal{L}_\text{task} = \sum_{(u,v,y)\in\mathcal{T}} (y\log \hat{y} + (1-y)\log(1-\hat{y})),
\end{align}
where $\mathcal{T}$ denotes the training set.

The overall loss of the proposed method is defined by:
\begin{align}
 	\label{eq:loss_all}
 	\mathcal{L}_\text{PERSCEN} = \mathcal{L}_\text{task} + \mathcal{L}_\text{VQ}. 
\end{align}
The training procedure of PERSCEN is summarized in Algorithm~\ref{algorithm}.
\begin{algorithm}
	\caption{The training process of PERSCEN}\label{algorithm}
	\While{not converge}{
		sample a mini-batch from all-scenario training data; \\
		\textbf{\textit{\# User-Specific Feature Graph Modeling}} \\
		obtain feature embeddings $\mathbf{e}_{u,1},\dots,\mathbf{e}_{u,N_f}$ by~(\ref{eq:emb-layer}); \\
		generate $\mathbf{A}^{(1)}_{u}$ of the first GNN layer by~(\ref{eq:generator}); \\
		generate high-order topology $\mathbf{A}^{(l)}_{u}$ by matrix product; \\
		generate $\bar{\mathbf{A}}^{(l)}_{u}$ for the $l$th GNN layer by refinement; \\
		obtain feature representations $\mathbf{h}_{u,1}^{(l)}, \cdots, \mathbf{h}_{u,N_f}^{(l)}$ by~(\ref{eq:gnn-update});\\
		obtain user hidden representations $\mathbf{h}_{u}^{(l)}$ by~(\ref{eq:user_hidden});\\
		\textbf{\textit{\# Scenario-Aware Preference Recognition}} \\
		obtain latent representation $\mathbf{z}_{u,s}$  by~(\ref{eq:scenario-sq});\\ 
		and scenario-aware preference codevector $\mathbf{c}_j$ by~(\ref{eq:vq});\\
		obtain scenario-aware preference representations $\mathbf{p}_{u,s}$ by~(\ref{eq:p_dyn}); \\
		obtain scenario context representations $\mathbf{p}_{s}$ by~(\ref{eq:p_sta}); \\
		obtain final scenario-aware preference $\hat{\mathbf{p}}_{u,s}$ by~(\ref{eq:personalized_scenario_pref}); \\
		\textbf{\textit{\# Progressive Scenario Adaptive Transfer}} \\
		\For{layer $l$}{
			obtain $l$th scenario-aware GLU output $\mathbf{g}^{(l)}_{u,s}$ by~(\ref{eq:asgu}); \\
		}
		obtain scenario-aware user representation $\hat{\mathbf{e}}_u$ by \eqref{eq:final_fusion2};\\ 
		obtain scenario-aware item representation $\hat{\mathbf{e}}_v$ by \eqref{eq:final_fusion2};\\
		calculate their matching score $\hat{y}$ by \eqref{eq:score};\\
		calculate task loss $\mathcal{L}_\text{task}$ by \eqref{eq:loss_task}, VQ loss $\mathcal{L}_\text{VQ}$ by \eqref{eq:loss_vq} ;\\
		calculate the overall loss of PERSCEN $\mathcal{L}_\text{PERSCEN}$ by \eqref{eq:loss_all}; \\
		take the gradient and update parameters.
	}
\end{algorithm}

\subsection{Comparison with Existing Works}

While prior works~\cite{xie2020internal,jiang2022adaptive,zhang2022scenario,zhao2023m5} have explored multi-scenario matching, PERSCEN introduces significant advancements. 
PERSCEN is the first model to incorporate user-specific modeling directly into the multi-scenario matching problem. 
In contrast, existing methods typically rely on fixed feature interaction patterns and generic scenario context features to learn preferences shared across all scenarios and scenario-aware preferences. 
PERSCEN overcomes these limitations by introducing a user-specific feature graph and utilizing vector quantization, 
facilitating personalized preference learning for both shared and scenario-aware aspects.
Extensive experimental results show PERSCEN's superior performance (Section~\ref{sec-perf}). Furthermore, a detailed efficiency analysis confirms that PERSCEN effectively balances performance with computational cost, ensuring its practical viability for real-world industrial systems (Section~\ref{sec-effi}).
Additionally, visualizations of feature graphs and scenario-aware preferences provide clear evidence that PERSCEN successfully captures user-specific and scenario-aware information for different users (Section~\ref{sec:case}).

\section{Experiments}
\subsection{Experiments Settings}
\subsubsection{Datasets}\label{sec:dataset}
We use two public benchmark datasets (Table~\ref{tab:dataset}): 
\begin{table}[!htbp]
	\centering
	\caption{Statistics of datasets used in this paper. }
	\setlength\tabcolsep{3pt}
	\begin{tabular}{ccccc}
		\toprule
		Dataset &\#Scenarios & \#Users & \#Items &  \#Interactions  \\
		\midrule
		KuaiRand-Pure & 4 & 24,122 & 7,583 & 650,283 \\
		Alimama & 4 & 138,622 & 846,811 & 933,633 \\
		\bottomrule
	\end{tabular}
	\label{tab:dataset}
	\vspace{-10pt}
\end{table}
\begin{itemize}[leftmargin=*]
	\item (i) \textbf{KuaiRand-Pure}~\cite{gao2022kuairand}: This dataset is collected from the recommendation logs of Kuaishou, a short-video platform. It consists of 30 days of interaction logs from 2022-04-08 to 2022-05-08. We split the data as follows: 2022-04-08 to 2022-04-21 for training, 2022-04-22 to 2022-04-28 for validation, and 2022-04-29 to 2022-05-08 for testing. We take the attribute tab as the scenario context. Since the top four scenarios account for more than 96\% of the total data volume, we select the four largest scenarios, labeled from K1 to K4, arranged in descending order of data volume. 
	\item (ii) \textbf{Alimama}~\cite{gai2017learning,zhou2018deep}: This dataset is released by Alimama, an online advertising platform. Data from 2017-05-06 to 2017-05-11 is used for training, 2017-05-12 for validation, and 2017-05-13 for testing. The dataset is divided into 4 scenarios based on city level, named from A1 to A4 in descending order of data volume. 
\end{itemize}
For both datasets, we focus on positive samples for training, i.e., clicked items. 
We also exclude users with fewer than 2 interactions to ensure a more robust training dataset. The maximum length of the user’s historical behavior sequence is set to 50. 
More implementation details are provided in Appendix~\ref{app:implementation_details}.
 
\subsubsection{Evaluation Metrics}
Following existing works~\cite{jiang2022adaptive,zhang2022scenario}, 
we evaluate the matching performance using Recall@K and Hits@K, 
Recall@K quantifies the proportion of relevant items within the top-$K$ retrieved candidates, 
and Hits@K indicates the presence of at least one relevant item among them.
In our experiments, we set the values of $K$ based on approximately 1\% of the total number of candidate items in the dataset. 
For KuaiRand-Pure, $K \in \{50, 100\}$, 
while for Alimama, $K \in \{500, 1000\}$. 

\subsection{Performance Comparison}\label{sec-perf}

We compare our PERSCEN
\footnote{Our code is available at \url{https://github.com/LARS-research/PERSCEN}.}
with two groups of baselines.
The first group is the single-scenario two-tower model, 
which only uses data from a single scenario for training,
marked with the suffix "-S":  
\textbf{YoutubeDNN-S}~\cite{davidson2010youtube} 
and 
\textbf{DSSM-S}~\cite{huang2013learning}. 
The second group is the multi-scenario two-tower model, which can be trained with all multi-scenario data: 
 \textbf{YoutubeDNN-M} (the suffix "-M" means multi-scenario data are used), \textbf{DSSM-M}, 
 \textbf{ICAN}~\cite{xie2020internal}, \textbf{ADIN}~\cite{jiang2022adaptive}, \textbf{SASS}~\cite{zhang2022scenario}, and \textbf{M5}~\cite{zhao2023m5}. 
We reproduce the baseline methods based on the publicly available code or the implementation details from the papers.
All results are averaged over five runs and are obtained on a 24GB NVIDIA GeForce RTX 3090Ti GPU. 
See Appendix~\ref{app:baselines}  for a detailed description.

Table~\ref{tab:results} shows the matching results. 
As shown, 
our PERSCEN outperforms all baselines in scenario-wise performance on the two datasets,
validating the effectiveness of user-specific modeling of multi-scenario information.
For scenarios with sufficient data, 
most multi-scenario matching models perform significantly better than single-scenario models, 
demonstrating the clear advantage of multi-scenario models in jointly modeling data from different scenarios. 
However, for relatively sparse data scenarios, the performance of models like ICAN and ADIN is worse than that of single-scenario models, 
as their optimization objectives are dominated by data-rich 
scenarios.
\begin{table}[h]
	\centering
	\caption{Efficiency and complexity comparison.}
	\label{tab:efficiency}
	\begin{tabular}{c|c|c|c}
		\hline
		\multirow{2}{*}{Model} & \multicolumn{3}{c}{\textit{KuaiRand-Pure}} \\ \cline{2-4} 
		& Training time (s) & Params (MB) & GFLOPs \\ \hline
		ADIN & \ms{8352}{1082} & 7.70 & 2.82  \\ \hline
		SASS & \ms{8745}{1028} & 3.27 & 10.19 \\ \hline
		M5 & \ms{10080}{1879} & 13.59 & 43.87 \\ \hline
		\textbf{PERSCEN} & \ms{13431}{2174} & 4.30 & 8.52  \\ \hline
		\hline
		\multirow{2}{*}{Model} & \multicolumn{3}{c}{\textit{Alimama}} \\ \cline{2-4} 
		& Training time (s) & Params (MB) & GFLOPs \\ \hline
		ADIN & \ms{5638}{355} & 39.56 & 4.22 \\ \hline 
		SASS & \ms{3977}{179} & 40.10 & 10.89 \\ \hline
		M5 & \ms{3861}{180} & 48.33 & 53.17 \\ \hline
		\textbf{PERSCEN} & \ms{6994}{265} & 39.80 & 12.84 \\ \hline
	\end{tabular}
\end{table}
\begin{table}[htbp]
	\centering
	\caption{Average inference time per batch with size 4096.}
	\begin{tabular}{cc}
		\toprule
		\textit{Alimama} & Average Inference Time (ms) \\
		\midrule
		ICAN & 3.19  \\
		ADIN & 9.23   \\
		SASS & 4.36    \\
		M5  & 4.59    \\
		\midrule
		PERSCEN w/ GRU & 5.02    \\
		\textbf{PERSCEN} & 4.83   \\
		\bottomrule
	\end{tabular}%
	\label{tab:infer}%
	\vspace{-10px}
\end{table}
In contrast, 
PERSCEN achieves impressive performance in all scenarios.
In particular, 
our method shows a significant performance improvement in scenarios with sparse data, 
which is attributed to shared codebook in scenario-aware preference recognition we introduce.
Preferences learned in data-rich scenarios can naturally transfer to data-sparse ones,
thus avoiding the issue of insufficient learning caused by sparse samples.
Experimental results in terms of hit rate demonstrate a similar phenomenon, 
and more experimental results are reported in Appendix~\ref{app:hits_results}.

\subsection{Efficiency Analysis}\label{sec-effi}

In terms of system complexity, 
we evaluated PERSCEN alongside mainstream baselines based on training time, model parameter size, and floating-point operations (FLOPs).
\begin{table*}[htbp]
	\caption{Test performance obtained on KuaiRand-Pure and Alimama. 
		The best results are bolded, the second-best results are underlined. 
		The proportion of each scenario’s data in the overall dataset is shown in $(\cdot)$ after the scenario identifier, highlighting data sparsity. For example, K1 (84\%) indicates that K1 data accounts for 84\% of the KuaiRand-Pure dataset.
	}
	\label{tab:results}
		\begin{tabular}{c|cc|cc|cc|cc}
			\hline
			\multirow{2}{*}{\textit{KuaiRand-Pure}} & \multicolumn{2}{c|}{K1~(84\%)} & \multicolumn{2}{c|}{K2~(9\%)} & \multicolumn{2}{c|}{K3~(4\%)} & \multicolumn{2}{c}{K4~(3\%)} \\	
			& R@50(\%) & R@100(\%) & R@50(\%) & R@100(\%) & R@50(\%) & R@100(\%) & R@50(\%) & R@100(\%)\\	
			\hline
			YoutubeDNN-S & \ms{16.66}{0.32} & \ms{27.61}{0.43} & \ms{17.30}{0.83} & \ms{25.48}{0.29} & \ms{11.44}{1.30} & \ms{17.60}{0.28} & \ms{8.21}{2.05} & \ms{12.72}{1.61}\\
			DSSM-S & \ms{17.71}{0.19} & \ms{28.42}{0.20} & \ms{20.04}{1.08} & \ms{29.15}{1.48} & \ms{12.55}{0.91} & \ms{19.00}{0.52} & \ms{8.34}{1.44} & \ms{13.45}{0.93}\\
			\hline
			YoutubeDNN-M & \ms{17.10}{0.18} & \ms{28.41}{0.20} & \ms{18.46}{0.53} & \ms{30.00}{0.42} & \ms{16.82}{0.48} & \ms{28.12}{0.42} & \ms{4.27}{0.47} & \ms{7.72}{0.52}\\
			DSSM-M & \ms{17.89}{0.17} & \ms{28.83}{0.12} & \ms{21.66}{0.53} & \ms{31.94}{0.60} & \ms{16.93}{0.68} & \ms{27.17}{0.58} & \ms{3.19}{0.30} & \ms{6.87}{0.46}\\
			ICAN & \ms{18.06}{0.16} & \ms{28.93}{0.15} & \ms{21.63}{0.62} & \ms{31.99}{0.37} & \ms{17.64}{0.29} & \ms{28.19}{0.61} & \ms{3.47}{0.26} & \ms{6.98}{0.32}\\
			ADIN & \sbms{18.46}{0.03} & \sbms{29.29}{0.25} & \ms{29.50}{0.52} & \ms{40.83}{0.39} & \sbms{18.88}{0.62} & \sbms{29.82}{1.37} & \sbms{19.12}{1.47} & \sbms{29.72}{1.71}\\
			SASS & \ms{17.70}{0.18} & \ms{28.28}{0.30} & \sbms{30.10}{0.31} & \sbms{41.25}{0.43} & \ms{17.99}{0.67} & \ms{28.03}{0.99} & \ms{18.95}{0.78} & \ms{28.17}{1.07}\\
			M5 & \ms{17.91}{0.08} & \ms{28.62}{0.25} & \ms{28.06}{0.47} & \ms{39.55}{0.42} & \ms{18.57}{0.95} & \ms{27.98}{0.65} & \ms{15.81}{1.55} & \ms{24.51}{1.27} \\	
			\TheName{} & \bms{18.74}{0.13} & \bms{29.95}{0.12} & \bms{30.69}{0.21} & \bms{42.62}{0.30} & \bms{19.60}{0.86} & \bms{30.50}{0.89} & \bms{21.39}{0.83} & \bms{31.68}{1.11} \\
			\hline \hline
			\multirow{2}{*}{\textit{Alimama}} & \multicolumn{2}{c|}{A1~(45\%)} & \multicolumn{2}{c|}{A2~(25\%)} & \multicolumn{2}{c|}{A3~(20\%)} & \multicolumn{2}{c}{A4~(10\%)} \\	
			& R@500(\%) & R@1000(\%) & R@500(\%) & R@1000(\%) & R@500(\%) & R@1000(\%) & R@500(\%) & R@1000(\%)\\	
			\hline
			YoutubeDNN-S & \ms{8.11}{1.01} & \ms{11.96}{0.74} & \ms{6.23}{0.88} & \ms{8.92}{0.59} & \ms{5.15}{1.48} & \ms{7.07}{1.24} & \ms{4.28}{1.09} & \ms{5.87}{1.12}\\
			DSSM-S & \ms{8.50}{1.23} & \ms{12.98}{0.86} & \ms{7.10}{1.01} & \ms{10.78}{0.61} & \ms{5.96}{1.95} & \ms{9.92}{1.64} & \ms{4.56}{1.61} & \ms{6.49}{1.33}\\
			\hline
			YoutubeDNN-M & \ms{9.47}{1.14} & \ms{14.22}{1.10} & \ms{8.83}{0.75} & \ms{12.98}{0.91} & \ms{8.91}{1.20} & \ms{12.88}{1.35} & \ms{9.09}{0.82} & \ms{12.76}{1.17}\\
			DSSM-M & \ms{9.58}{1.27} & \ms{14.43}{1.47} & \ms{9.49}{1.20} & \ms{14.46}{1.32} & \ms{9.34}{1.41} & \ms{14.03}{1.66} & \ms{9.93}{1.25} & \ms{14.94}{1.42} \\
			ICAN & \ms{9.76}{1.35} & \ms{15.02}{1.08} & \ms{9.81}{1.16} & \ms{15.28}{0.90} & \ms{9.51}{1.24} & \ms{14.91}{0.98} & \ms{10.18}{1.27} & \ms{15.55}{0.90}\\
			ADIN & \ms{11.01}{0.52} & \ms{15.50}{0.62} & \sbms{11.38}{0.67} & \sbms{16.13}{0.71} & \sbms{10.94}{0.34} & \sbms{16.02}{0.50} & \sbms{11.82}{0.50} & \sbms{16.72}{0.38} \\
			SASS & \ms{10.39}{0.14} & \ms{15.06}{0.10} & \ms{10.80}{0.17} & \ms{15.72}{0.20} & \ms{10.36}{0.24} & \ms{15.32}{0.31} & \ms{11.04}{0.58} & \ms{15.86}{0.44} \\
			M5 & \sbms{11.01}{0.36} & \sbms{15.71}{0.27} & \ms{11.06}{0.17} & \ms{15.56}{0.19} & \ms{10.95}{0.26} & \ms{15.49}{0.44} & \ms{11.00}{0.60} & \ms{15.45}{0.78} \\	
			\TheName{} & \bms{12.72}{0.20} & \bms{17.63}{0.37} & \bms{12.66}{0.32} & \bms{17.60}{0.22} & \bms{12.34}{0.24} & \bms{17.32}{0.14} & \bms{12.63}{0.12} & \bms{17.47}{0.42} \\
			\hline
		\end{tabular}
\end{table*}
\begin{table*}[htbp]
	\caption{Ablation study on Alimama.}
	\setlength\tabcolsep{3pt}
	\label{tab:ablation}
	\begin{tabular}{c|c|cc|cc|cc|cc}
		\hline
		\multirow{2}{*}{Type}&\multirow{2}{*}{Variant} & \multicolumn{2}{c|}{A1~(45\%)} & \multicolumn{2}{c|}{A2~(25\%)} & \multicolumn{2}{c|}{A3~(20\%)} & \multicolumn{2}{c}{A4~(10\%)} \\	
		&& R@500(\%) & R@1000(\%) & R@500(\%) & R@1000(\%) & R@500(\%) & R@1000(\%) & R@500(\%) & R@1000(\%)\\	
		\hline
		\multirow{2}{*}{\makecell{Feature graph\\ modeling}}&w/o GNN & \ms{12.10}{0.50} & \ms{17.02}{0.86} & \ms{12.17}{0.38} & \ms{16.99}{0.43} & \ms{11.94}{0.47} & \ms{16.77}{0.64} & \ms{12.33}{0.38} & \ms{17.05}{0.85}\\
		&w/ shared graph & \ms{12.35}{0.27} & \ms{17.41}{0.67} & \ms{12.33}{0.34} & \ms{17.46}{0.47} & \ms{12.03}{0.41} & \ms{17.03}{0.41} & \ms{12.48}{0.49} & \ms{17.38}{0.46} \\
		\hline
		\multirow{2}{*}{\makecell{Preference \\recognition}}&w/o spec sequence & \ms{11.79}{0.26} & \ms{16.48}{0.39} & \ms{11.81}{0.31} & \ms{16.56}{0.38} & \ms{11.46}{0.34} & \ms{16.44}{0.36} & \ms{11.99}{0.40} & \ms{16.66}{0.53} \\
		&w/o VQ & \ms{11.98}{0.33} & \ms{16.70}{0.58} & \ms{12.00}{0.34} & \ms{16.70}{0.41} & \ms{11.71}{0.30} & \ms{16.59}{0.37} & \ms{12.50}{0.45} & \ms{17.07}{0.39} \\
		\hline
		Scenario transfer&w/o GLU & \ms{11.82}{0.13} & \ms{16.42}{0.20} & \ms{11.96}{0.21} & \ms{16.44}{0.33} & \ms{11.74}{0.32} & \ms{16.31}{0.33} & \ms{11.96}{0.24} & \ms{16.50}{0.25} \\
		\hline
		&\TheName{} & \bms{12.72}{0.20} & \bms{17.63}{0.37} & \bms{12.66}{0.32} & \bms{17.60}{0.22} & \bms{12.34}{0.24} & \bms{17.32}{0.14} & \bms{12.63}{0.12} & \bms{17.47}{0.42} \\
		\hline
	\end{tabular}
\end{table*}
As shown in Table~\ref{tab:efficiency}, 
the increase in system complexity across multiple dimensions for PERSCEN is minimal, confirming that our design effectively balances both performance and efficiency. 
Notably, 
the FLOPs of PERSCEN (\textasciitilde10 GFLOPs) fall well within the deployment range of industrial recommendation systems.

Inference efficiency is also crucial for matching models. Thus, we further compare the inference time of PERSCEN with other baselines on the Alimama test set. The experiments were conducted on an NVIDIA GeForce RTX 3090Ti GPU with a batch size of 4096. 
Table~\ref{tab:infer} shows the average inference time per batch. 
The results indicate that PERSCEN’s inference time slightly increases compared to the baselines due to user-specific feature interaction modeling and scenario preference recognition. However, this increase is minor. 
ICAN relys on a simple scenario-aware attention mechanism for multi-scenario modeling to achieve the lowest inference time, but its performance is unsatisfactory.
Notably, compared to the variant using GRU, PERSCEN achieves a lower average inference time, highlighting that the scenario-aware GLU offers more efficient transfer capabilities.

\subsection{Ablation Study}

We further compare PERSCEN with the following variants:
\begin{itemize}
	\item \textbf{w/o GNN} 
	foregoes the GNN for capturing higher-order feature interaction patterns on the user/item side, 
	opting instead for a simpler MLP network for feature crossing.
	\item \textbf{w/ shared graph} employs a single, shared adjacency matrices for all user/items, 
	unlike PERSCEN, 
	which learns user-specific adjacency matrices; 
	\item \textbf{w/o spec sequence} does not utilize scenario-specific behavior sequences to distil user-specific and scenario-aware preferences, 
	but instead only uses scenario context feature to learn general scenario-aware preferences; 
	\item \textbf{w/o VQ} removes the VQ mechanism, directly fusing the scenario context representation $p_s$ with the scenario-specific sequence embedding $z_{u,s}$;
	\item \textbf{w/o GLU} replaces our scenario-aware GLU with a direct fusion of the final GNN output $h_u^L$ and the scenario-aware preference representation $\hat{p}_{u,s}$.
\end{itemize}

Table~\ref{tab:ablation} presents the results.
As shown,
"w/o GNN" significantly underperforms PERSCEN, 
demonstrating that using a GNN to model feature interactions is more expressive and effective than a simple MLP. 
The performance gain of PERSCEN over "w/ shared graph" highlights the importance of learning user-specific feature interaction patterns for multi-scenario matching tasks.  
We can also observe that "w/o spec sequence" performs worse than PERSCEN, 
which validates the necessity of learning user-specific and scenario-aware preferences based on scenario-specific sequences.
Particularly in data-sparse scenarios, 
\begin{figure*}[htbp]
	\centering
	\includegraphics[width=1\textwidth]{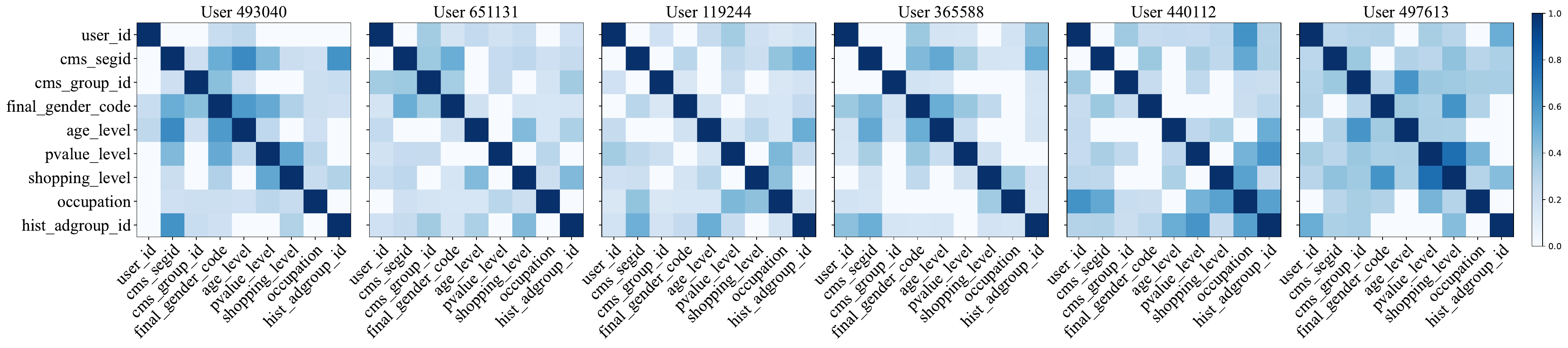}
	\caption{Visualizations of user-specific adjacency matrices generated by PERSCEN.
	}
	\label{fig:usergraph}
\end{figure*}
\begin{figure}[htbp]
	\centering
	\includegraphics[width=0.40\textwidth]{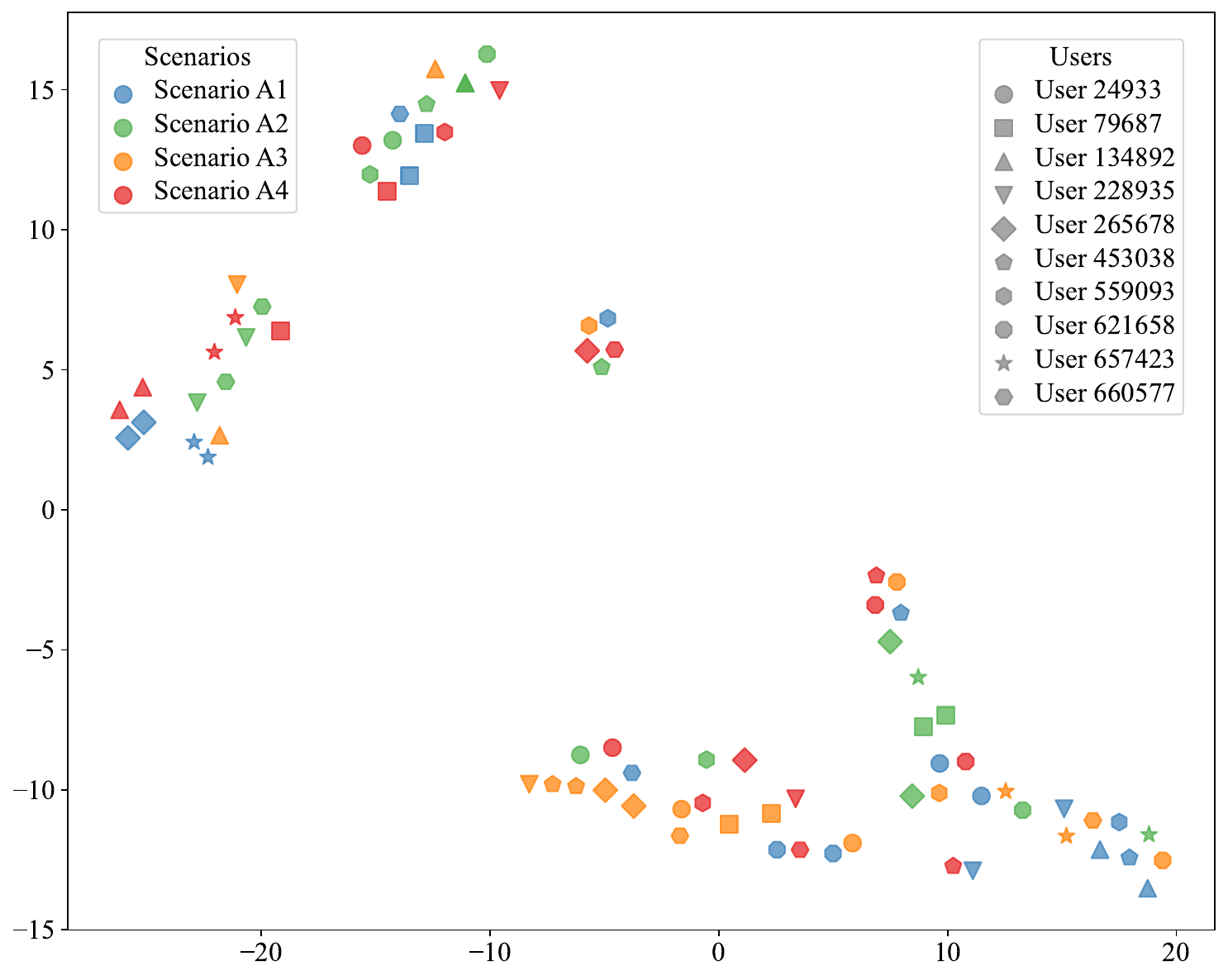}
	\caption{Visualizations of user-specific and scenario-aware preference generated by PERSCEN.
	}
	\vspace{-10pt}
	\label{fig:tsne}
\end{figure}
the codebook effectively mitigates multi-scenario data imbalance by learning shared scenario-aware preference vectors across different scenarios.

\subsection{Case Study}\label{sec:case}

\subsubsection{User-Specific Feature Interaction}
To validate the effectiveness of our method in learning user-specific feature interaction patterns, 
we randomly sample six users from the Alimama dataset and visualize their user-specific feature graph adjacency matrices in Figure~\ref{fig:usergraph}. 
The results clearly show that PERSCEN successfully learns unique feature graphs for each user.
For example, with user ID 493040, 
PERSCEN captures a strong interaction between between <cms\_segid> (content management system segment ID) and <age\_level>, 
suggesting that this user’s content preferences across scenarios are tightly linked to their age group. 
In contrast, for user ID 497613, 
the primary interaction involves consumption level and shopping frequency, 
highlighting a distinctly different set of behavioral patterns across scenarios for that individual.

\subsubsection{User-Specific and Scenario-Aware Preference}
To confirm the effectiveness of our method in learning user-specific and scenario-aware preferences,
we use t-SNE~\cite{van2008visualizing} to visualize the scenario-aware preference representations for 10 randomly sampled users across different scenarios in the Alimama dataset (Figure~\ref{fig:tsne}). 
In this visualization, 
distinct colors represent different scenarios, 
while different shapes denote individual users.
The results clearly demonstrate that PERSCEN effectively captures user-specific and scenario-aware intentions. Even within the same scenario, 
users exhibit distinct scenario-specific preferences, 
reflecting their unique historical interactions within that context. 
Interestingly, 
some users also display similar scenario-aware preferences across different scenarios.
For example, in the upper left corner of Figure~\ref{fig:tsne}, 
user ID 134892 in Scenario A4 (red triangle) and user ID 265678 in Scenario A1 (blue rhombus) show nearly identical representations,
suggesting that their scenario-aware preferences are very similar. 
This finding highlights a significant advantage of PERSCEN: 
its ability to learn scenario-aware preferences even in data-sparse scenarios. 
By allowing preference knowledge in the codebook to be shared across scenarios, 
PERSCEN avoids the limitations of methods that rely solely on scenario context, 
which often struggle when scenario-specific data is scarce.

\section{Conclusion}

In this study, 
we introduce PERSCEN, 
an efficient multi-scenario matching model specifically designed for user-specific modeling. 
PERSCEN achieves this through three core innovations: we builds a user-specific feature graph and uses a lightweight graph neural network to capture user-specific interaction patterns. 
To effectively model contextual differences, 
we employ a vector quantization-based mechanism that distils user-specific and scenario-aware preferences directly from behavior sequences. 
Finally, to efficiently combine these shared and scenario-specific preferences, 
we developed a progressive scenario-aware gated linear unit, facilitating adaptive and efficient information integration.
Extensive experiments demonstrate PERSCEN's superior performance compared to existing methods. 
Further efficiency analysis confirms that PERSCEN strikes an effective balance between performance and computational cost, ensuring its practical viability for real-world industrial systems.

\section{Acknowledgments}
We thank the anonymous reviewers for their valuable comments.
This research was supported by the National Natural Science Foundation of China (Nos. U22B2036, 62261136549), 
the Technological Innovation Team of Shaanxi Province (No. 2025RS-CXTD-009), 
the International Cooperation Project of Shaanxi Province (No. 2025GH-YBXM-017), 
the Fundamental Research Funds for the Central Universities (Nos. G2024WD0151, D5000240309), 
and the XPLORER PRIZE.
Q. Yao is sponsored by CCF-Zhipu Large Model Innovation Fund (No. Zhipu202402).
\clearpage
{
	\bibliographystyle{ACM-Reference-Format}
	\balance
	\bibliography{multi-scenario}
}
\clearpage
\appendix
\begin{table*}[htbp]
	\caption{Hyperparameters used by \TheName{}.}
	\centering
	\label{tab:hyperpara}
		\begin{tabular}{l|l|l|l}
			\hline
			Hyperparameter & Range  &KuaiRand-Pure& Alimama\\
			\hline	
			number of GNN layers & $[1, 2, 3, 4]$ & 3&1 \\
			codebook size & $[0, 2,5,10,20,50,100]$ &10 &5 \\
			$\beta$ in loss function & $[0, 0.25,\dots,1]$ & 0.25&0.25 \\
			embedding dimension &$[10,11,\dots,20]$& 16&16 \\ 
			batch size & $[1024, 2048, \cdots, 9192]$ & 4096&4096 \\
			learning rate & $[1e-4,1e-1]$ & 0.001&0.001 \\
			weight decay & $[1e-7,1e-5]$ & 1e-6&1e-6\\
			sequence pooling strategy &$[\text{sum},\text{max},\text{mean},\text{concat}]$ & mean & mean \\ 
			the number of negative sample for each positive sample &$[1,\cdots,10]$ &10 & 10 \\ \hline
		\end{tabular}
\end{table*}
\section{Implementation Details}
\label{app:implementation_details}

\subsection{Parameter Setting}
We use Adam optimizer~\cite{kingma2014adam}. 
To search for the appropriate hyperparameters, 
we divided the dataset into a validation set. The specific details of the division can be found in Section~\ref{sec:dataset}.
For the stage of retrieval, 
after getting the user and item embedding, 
we apply FAISS retrieval system to execute ANN seatch process.
The hyperparameters and their range used by \TheName{} are summarized in Table~\ref{tab:hyperpara}.

\subsection{Baselines}
\label{app:baselines}
We compare the proposed PERSCEN with the following two groups of baselines.
The first group is the single-scenario two-tower model, 
which only uses data from a single scenario  for training,
marked with the suffix "-S": 
\begin{itemize}[leftmargin=*]
	\item \textbf{YoutubeDNN-S}~\cite{davidson2010youtube} adopts average pooling to extract user’s interest with a sampled softmax loss to optimize similarities between users and items.
	\item \textbf{DSSM-S}~\cite{huang2013learning} builds a relevance score model to extract user and item representations with two-tower architecture.
\end{itemize}
The second group is the multi-scenario two-tower model, which can be trained with all multi-scenario data:  
\begin{itemize}[leftmargin=*]
	\item \textbf{YoutubeDNN-M} is a modified version of YoutubeDNN which is optimized with multi-scenario data during the training process, with the scenario indicator treated as a regular feature.
	\item \textbf{DSSM-M} is a modified version of DSSM which is similar with YoutubeDNN-M, 
	taking multi-scenario data as input.
	\item \textbf{ICAN}~\cite{xie2020internal} introduces a scenario-aware internal and contextualized attention mechanism to modulate the multi-scenario user behavior sequence, which is then passed into a shared backbone to produce the final prediction.  
	\item \textbf{ADIN}~\cite{jiang2022adaptive} uses a shared backbone but introduces separate scenario predictors to capture both the commonalities and distinctions between scenarios. 
	\item \textbf{SASS}~\cite{zhang2022scenario} adds a scenario encoder to capture scenario preferences and a multi-layer scenario-adaptive transfer module to guide the shared backbone, achieving improved performance. 
	\item \textbf{M5}~\cite{zhao2023m5} employs scenario context to modulate the behavior sequence and uses a Mixture of Experts to capture the commonalities across different scenarios via a shared backbone.
\end{itemize}

\section{More Experimental Results}
\label{app:results}

\subsection{Performance Comparison}
Table~\ref{tab:hits_results} presents the hit ratio performance. 
Similar to recall, 
PERSCEN surpasses existing methods in hit rate, 
confirming its effectiveness in user-specific modeling.

\label{app:hits_results}
\begin{table*}[htbp]
	\caption{Test performance obtained on KuaiRand-Pure and Alimama. 
		The best results are bolded, the second-best results are underlined. 
		The proportion of each scenario’s data in the overall dataset is shown in $(\cdot)$ after the scenario identifier, highlighting data sparsity. For example, K1 (84\%) indicates that K1 data accounts for 84\% of the KuaiRand-Pure dataset.
	}
	\label{tab:hits_results}
	\begin{tabular}{c|cc|cc|cc|cc}
		\hline
		\multirow{2}{*}{\textit{KuaiRand-Pure}} & \multicolumn{2}{c|}{K1~(84\%)} & \multicolumn{2}{c|}{K2~(9\%)} & \multicolumn{2}{c|}{K3~(4\%)} & \multicolumn{2}{c}{K4~(3\%)} \\	
		& H@50(\%) & H@100(\%) & H@50(\%) & H@100(\%) & H@50(\%) & H@100(\%) & H@50(\%) & H@100(\%)\\	
		\hline
		YoutubeDNN-M & \ms{39.30}{0.15} & \ms{55.10}{0.26} & \ms{25.27}{0.50} & \ms{39.01}{0.37} & \ms{30.73}{1.08} & \ms{46.24}{0.33} & \ms{5.55}{0.60} & \ms{9.52}{0.55}\\
		DSSM-M & \ms{40.45}{0.16} & \ms{55.67}{0.21} & \ms{28.88}{0.45} & \ms{40.90}{0.35} & \ms{31.00}{0.81} & \ms{45.19}{0.96} & \ms{4.86}{0.67} & \ms{8.85}{0.42}\\
		ICAN & \ms{40.73}{0.45} & \ms{55.84}{0.32} & \ms{29.11}{0.34} & \ms{41.11}{0.20} & \ms{30.61}{0.70} & \ms{45.53}{1.17} & \ms{4.76}{0.37} & \ms{8.66}{0.41}\\
		ADIN & \sbms{41.44}{0.15} & \sbms{56.38}{0.19} & \ms{38.27}{0.55} & \sbms{50.67}{0.42} & \sbms{34.33}{1.16} & \sbms{47.73}{1.66} & \ms{21.91}{1.78} & \sbms{33.74}{2.11}\\
		SASS & \ms{40.49}{0.30} & \ms{55.44}{0.26} & \sbms{38.56}{0.59} & \ms{50.57}{0.32} & \ms{33.67}{1.20} & \ms{46.85}{0.85} & \sbms{22.13}{1.26} & \ms{33.18}{1.35}\\
		M5 & \ms{40.73}{0.09} & \ms{55.63}{0.11} & \ms{36.32}{0.55} & \ms{48.57}{0.46} & \ms{32.88}{0.96} & \ms{46.05}{1.23} & \ms{17.37}{1.25} & \ms{26.80}{0.64}\\	
		\TheName{} & \bms{41.78}{0.30} & \bms{56.88}{0.13} & \bms{39.36}{0.26} & \bms{52.18}{0.36} & \bms{34.81}{0.94} & \bms{48.78}{0.63} & \bms{24.35}{1.75} & \bms{35.54}{0.64}\\
		\hline \hline
		\multirow{2}{*}{\textit{Alimama}} & \multicolumn{2}{c|}{A1~(45\%)} & \multicolumn{2}{c|}{A2~(25\%)} & \multicolumn{2}{c|}{A3~(20\%)} & \multicolumn{2}{c}{A4~(10\%)} \\	
		& H@500(\%) & H@1000(\%) & H@500(\%) & H@1000(\%) & H@500(\%) & H@1000(\%) & H@500(\%) & H@1000(\%)\\	
		\hline
		YoutubeDNN-M & \ms{14.87}{0.15} & \ms{20.64}{0.11} & \ms{13.24}{0.24} & \ms{20.45}{0.06} & \ms{13.97}{0.15} & \ms{20.05}{0.35} & \ms{14.55}{0.31} & \ms{21.26}{0.27}\\
		DSSM-M & \ms{14.12}{1.75} & \ms{20.51}{1.16} & \ms{13.92}{1.56} & \ms{20.71}{1.10} & \ms{13.78}{1.73} & \ms{20.36}{1.03} & \ms{14.91}{1.82} & \ms{21.13}{0.94}\\
		ICAN & \sbms{16.35}{0.43} & \sbms{22.41}{0.27} & \sbms{16.63}{0.49} & \ms{23.07}{0.28} & \ms{16.18}{0.54} & \ms{22.53}{0.56} & \ms{16.92}{0.57} & \ms{23.25}{0.34}\\
		ADIN & \ms{16.05}{0.73} & \ms{21.95}{0.79} & \ms{16.62}{0.95} & \sbms{23.10}{0.88} & \sbms{16.24}{0.53} & \sbms{23.02}{0.68} & \sbms{17.23}{0.71} & \sbms{23.78}{0.56}\\	
		SASS & \ms{15.50}{0.38} & \ms{21.71}{0.30} & \ms{15.88}{0.49} & \ms{22.55}{0.42} & \ms{15.68}{0.48} & \ms{22.16}{0.38} & \ms{16.48}{0.41} & \ms{22.67}{0.66}\\
		M5 & \ms{15.48}{1.30} & \ms{21.57}{1.21} & \ms{15.69}{0.70} & \ms{22.04}{0.98} & \ms{15.62}{1.05} & \ms{21.74}{1.03} & \ms{16.02}{0.65} & \ms{21.70}{0.96}\\
		\TheName{} & \bms{18.42}{0.24} & \bms{24.87}{0.45} & \bms{18.63}{0.44} & \bms{25.31}{0.31} & \bms{18.07}{0.38} & \bms{24.69}{0.15} & \bms{18.67}{0.27} & \bms{24.90}{0.50}\\
		\hline
	\end{tabular}
\end{table*}

\subsection{Hyperparameter Sensitivity Analysis}
\label{app:hyper}
\subsubsection{Number of GNN Layers}

\begin{figure}[h]
	\centering
	\subfigure[Scenario A1]{
		\includegraphics[width=0.225\textwidth]{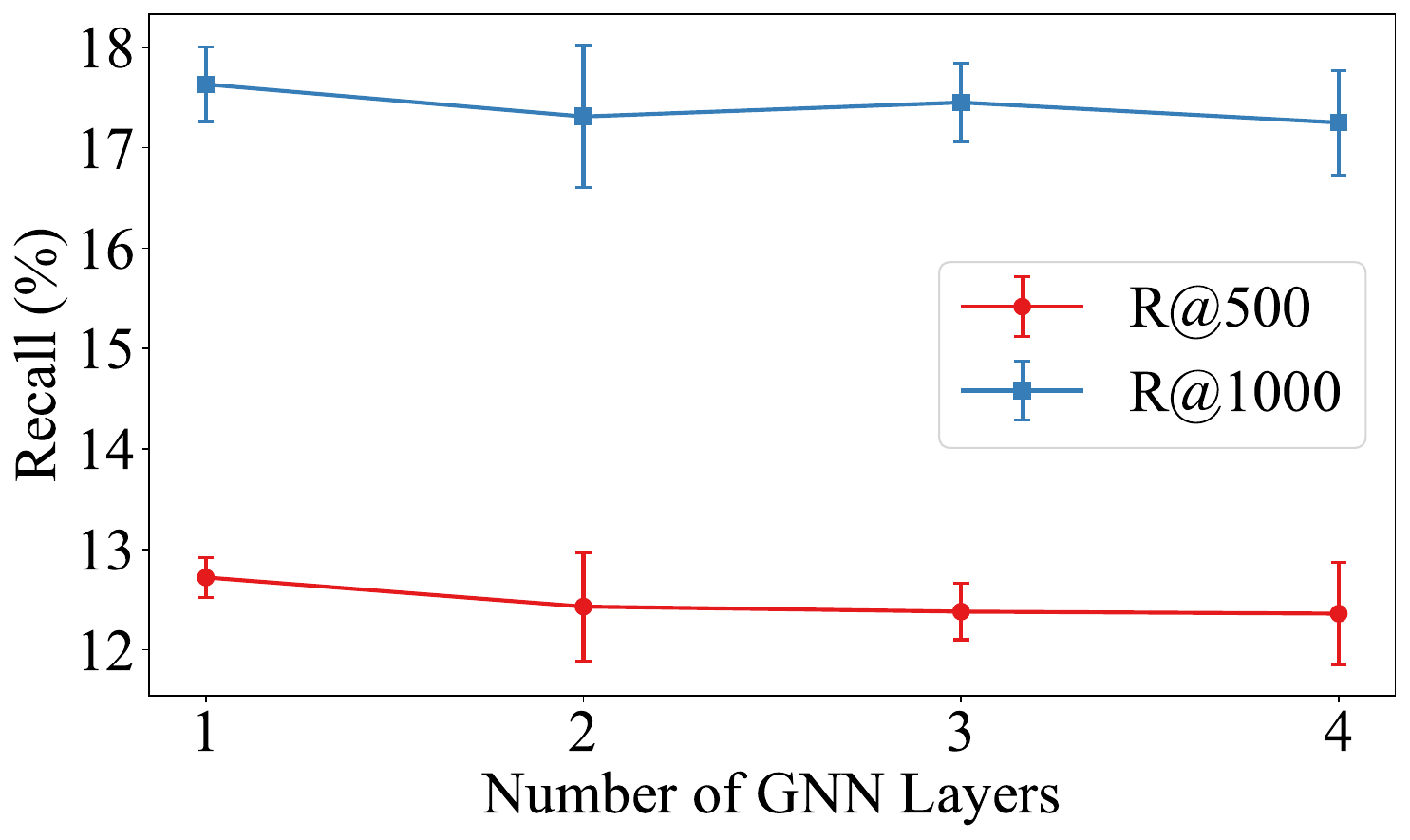}
	}
	\subfigure[Scenario A2]{
		\includegraphics[width=0.225\textwidth]{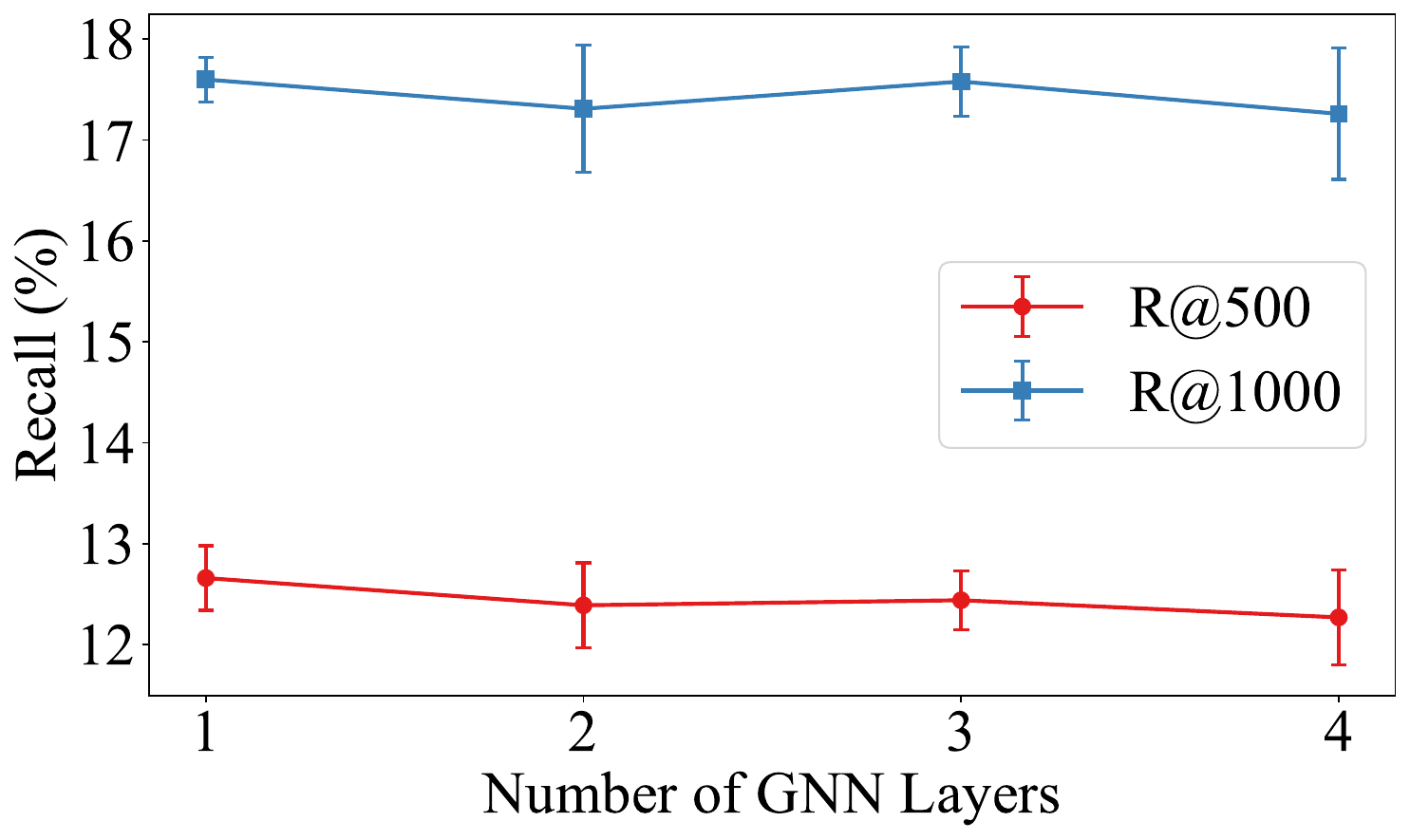}
	}
	\subfigure[Scenario A3]{
		\includegraphics[width=0.225\textwidth]{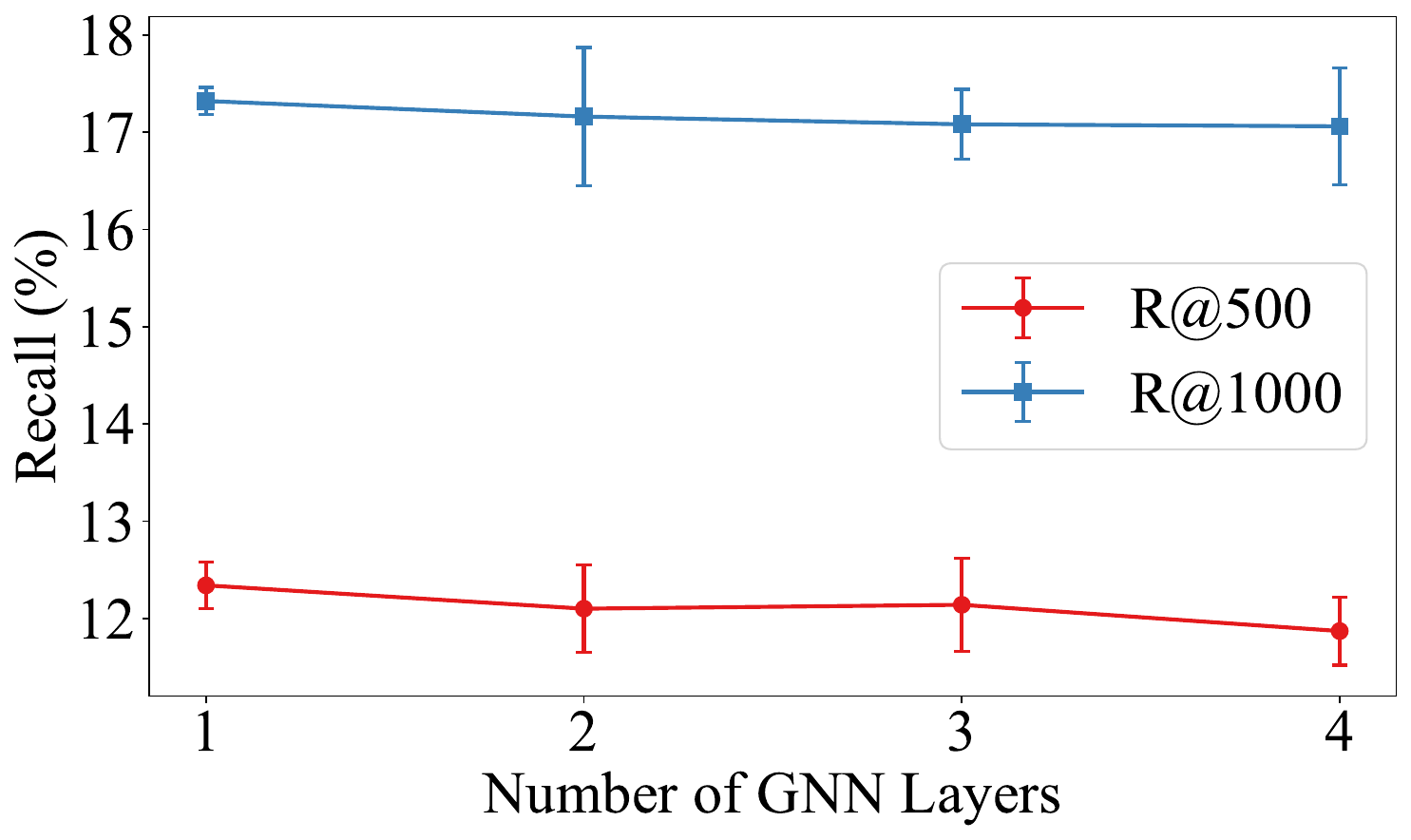}
	}
	\subfigure[Scenario A4]{
		\includegraphics[width=0.225\textwidth]{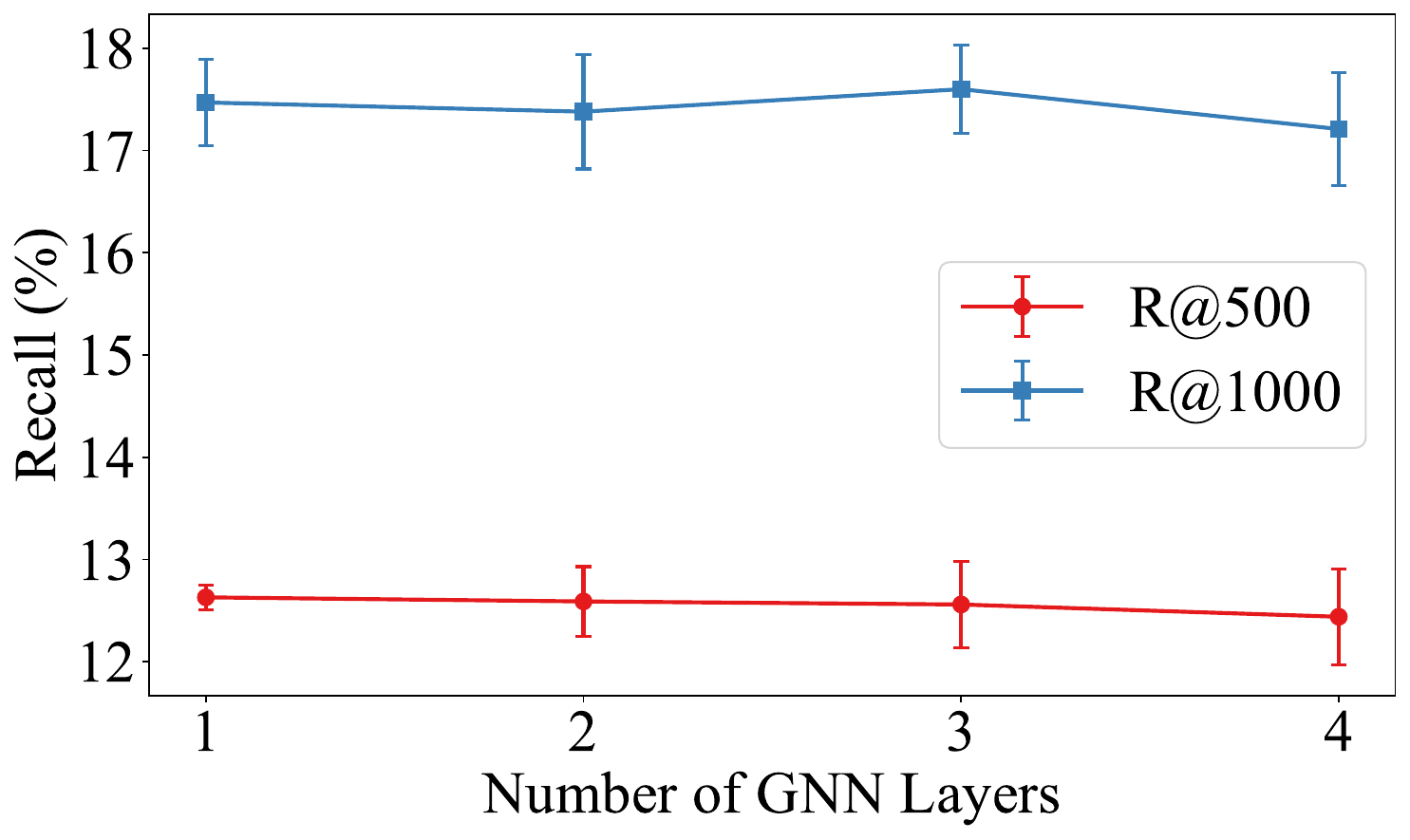}
	}
	\caption{Varying the number of GNN layers for PERSCEN in Alimama.}
	\label{fig:layeranalysis}
\end{figure}

In this section, 
we demonstrate the impact of the number of GNN layers on test performance across different scenarios in Alimama dataset,
a hyperparameter that is closely related to modeling of user-specific feature interactions. 
As shown in Figure~\ref{fig:layeranalysis}, 
it can be observed that for the Alimama dataset, which has fewer user feature fields (9), 
using just one GNN layer is sufficient to achieve optimal performance. 

\subsubsection{Codebook Size}
\begin{figure}[htbp]
	\centering
	\subfigure[Scenario A1]{
		\includegraphics[width=0.225\textwidth]{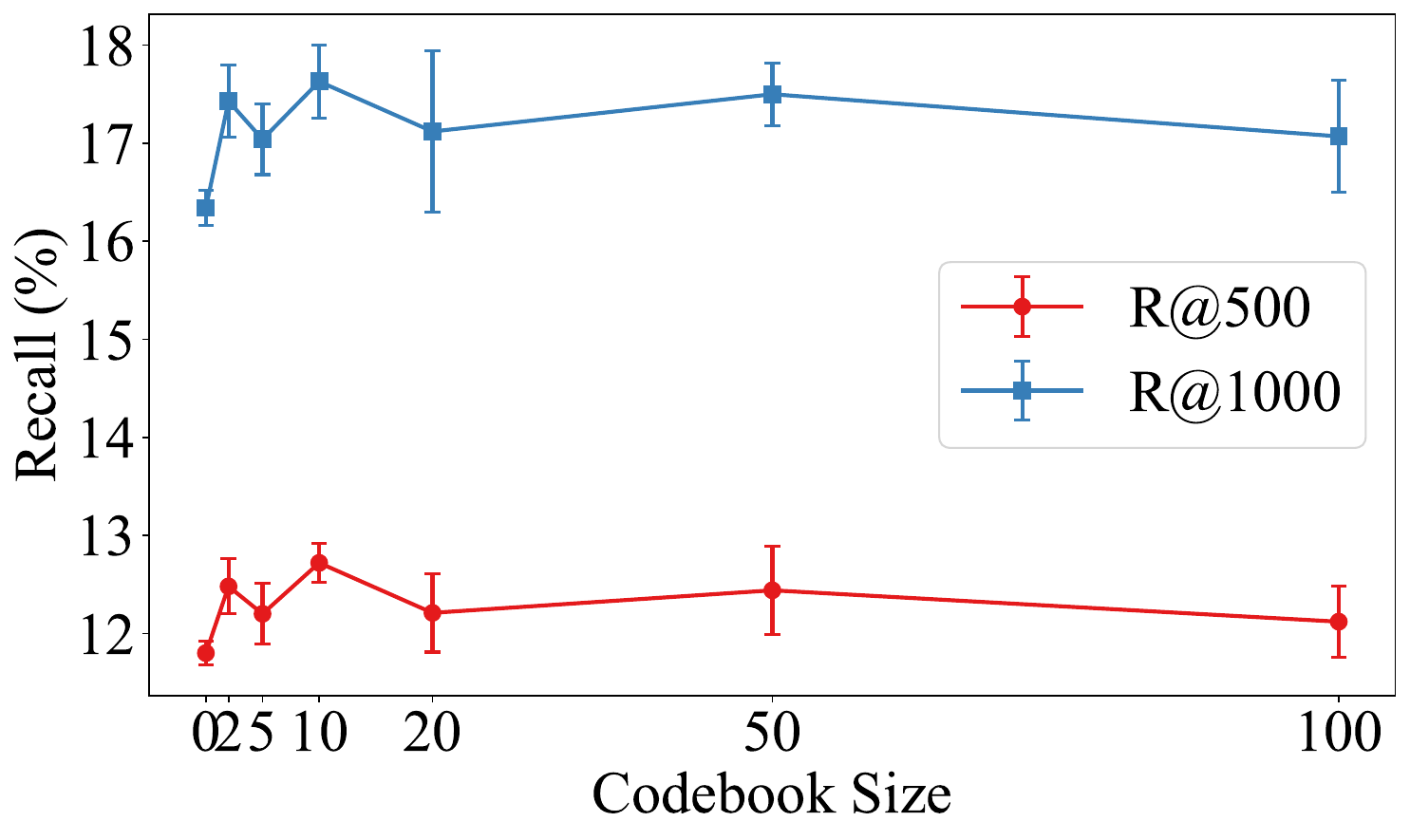}
	}
	\subfigure[Scenario A2]{
		\includegraphics[width=0.225\textwidth]{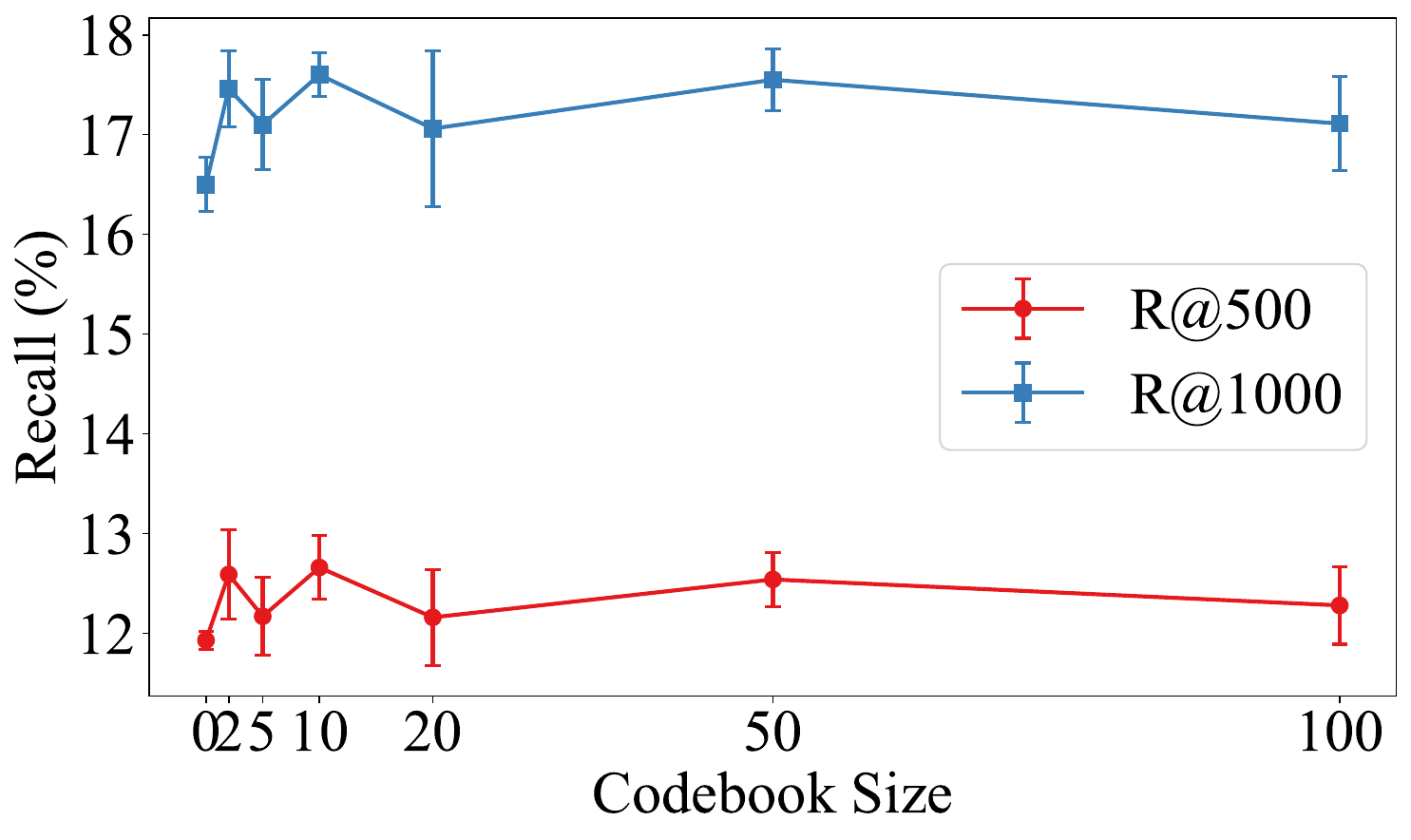}
	}
	\subfigure[Scenario A3]{
		\includegraphics[width=0.225\textwidth]{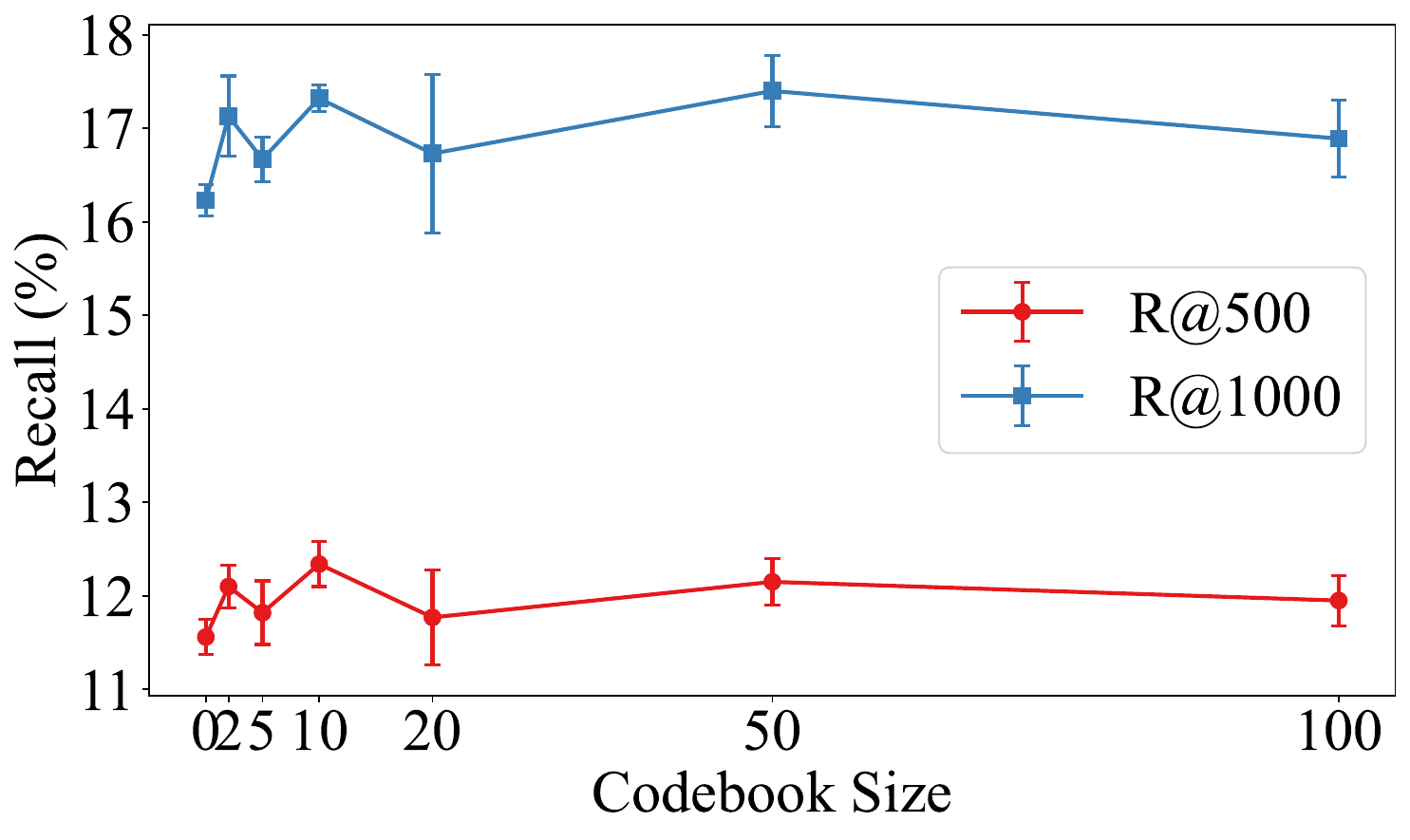}
	}
	\subfigure[Scenario A4]{
		\includegraphics[width=0.225\textwidth]{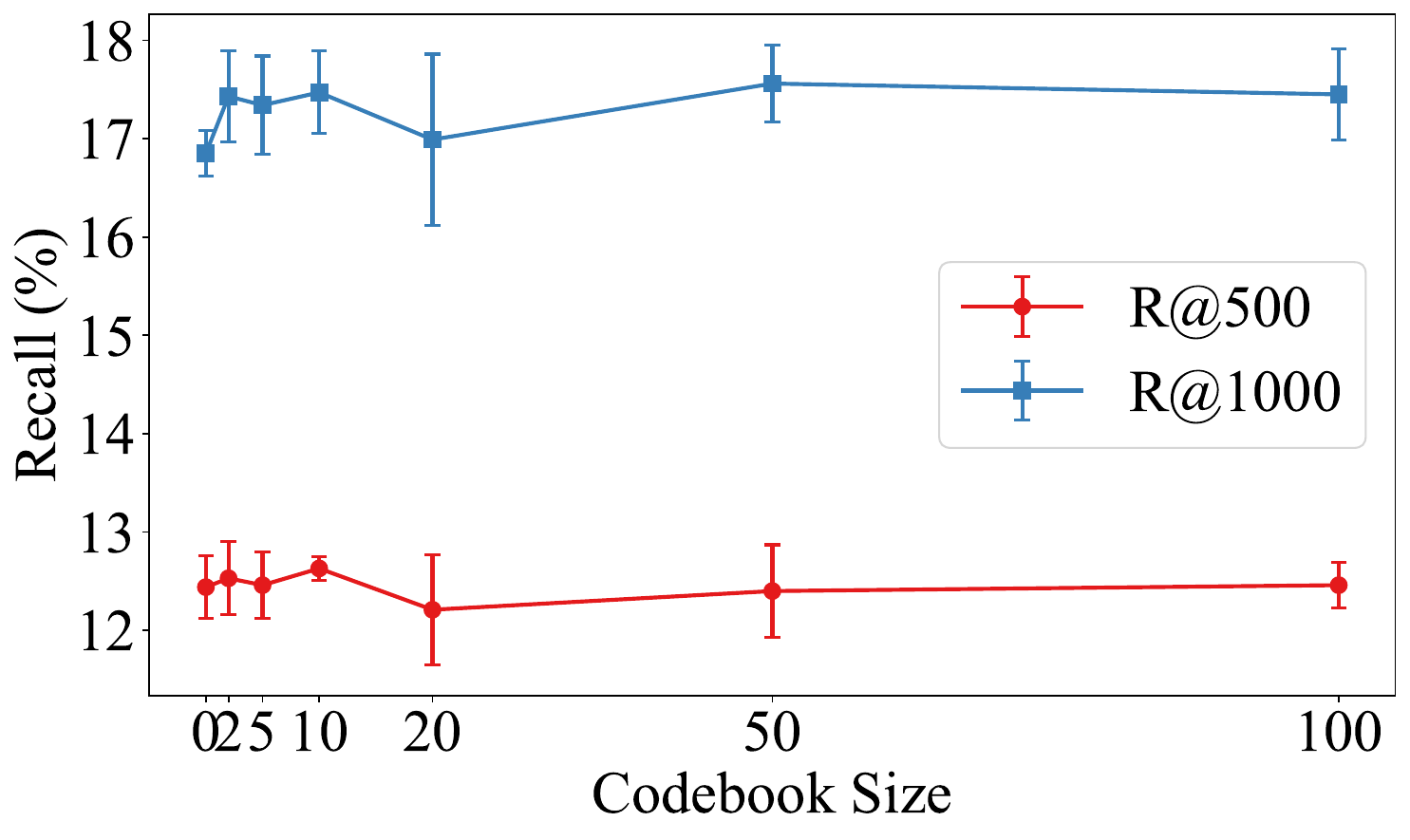}
	}
	\caption{Varying the codebook size for PERSCEN in Alimama.}
	\label{fig:codebooksizeanalysis}
\end{figure}
In this section, 
we evaluate the impact of codebook size on test performance across different scenarios in Alimama dataset,
which plays an important role in learning scenario-aware preferences. 
As shown in Figure~\ref{fig:codebooksizeanalysis}, 
when there is no codebook (codebook size = 0), 
i.e., when the latent representations $\mathbf{z}_{u,s}$ obtained from the scenario sequence are directly integrated with scenario representation without considering vector quantization techniques, 
there is a significant performance drop. 
This confirms the effectiveness of vector quantization (VQ) in learning user-specific and scenario-aware preferences, 
which can effectively improve the model’s performance across various scenarios. 
It is worth noting that a larger codebook size is not always better. 
The optimal codebook size for the Alimama dataset is 10. 
As the codebook size increases, performance first improves and then decreases. 
One possible reason is that as the codebook size grows, 
the boundaries between diverse scenario-aware preference clusters become blurred, 
leading to insufficient learning of individual scenario-aware preference code vectors. 
Therefore, the codebook size needs to be finely tuned to avoid negative effects. 

\begin{figure}[h]
	\centering
	\subfigure[Scenario A1]{
		\includegraphics[width=0.225\textwidth]{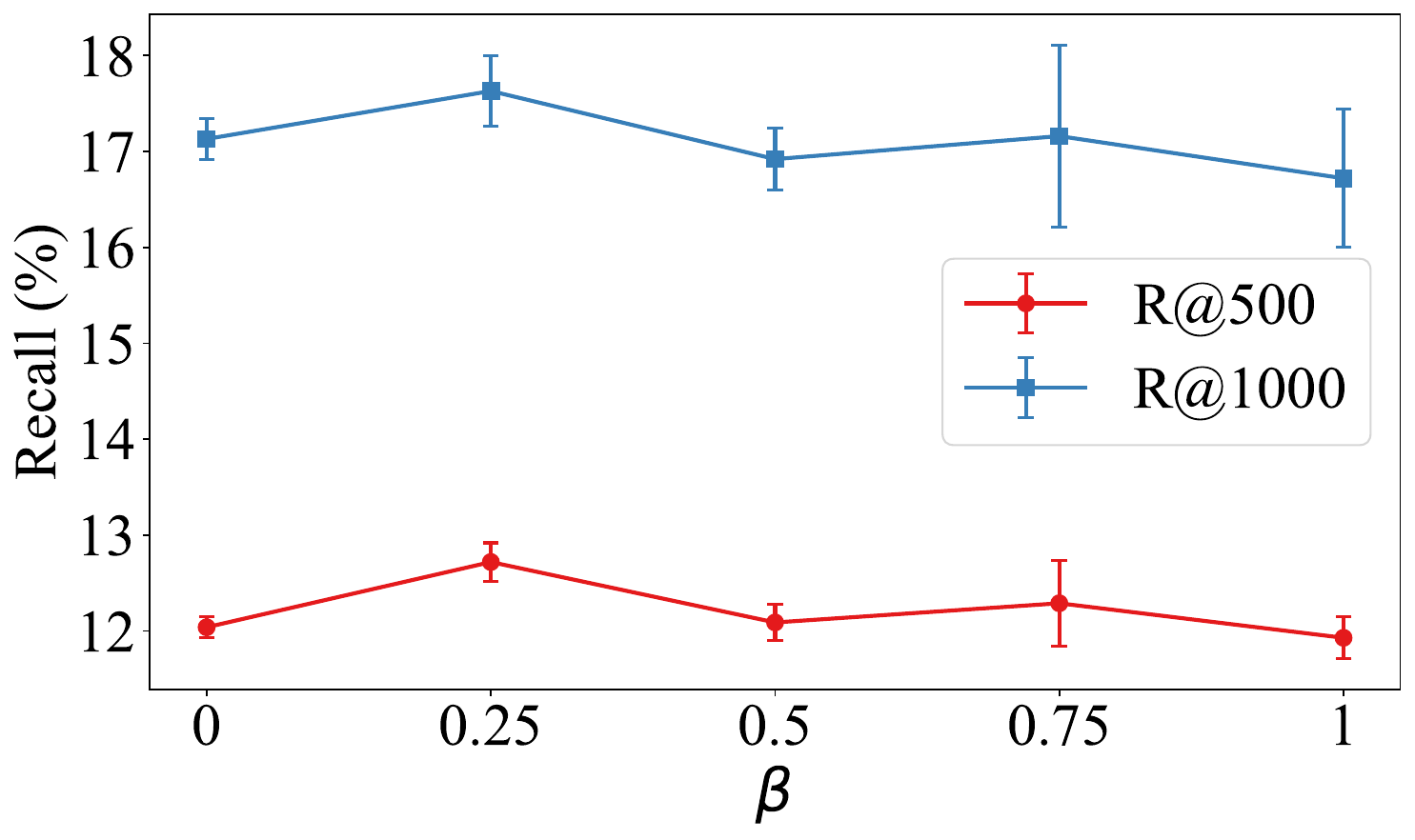}
	}
	\subfigure[Scenario A2]{
		\includegraphics[width=0.225\textwidth]{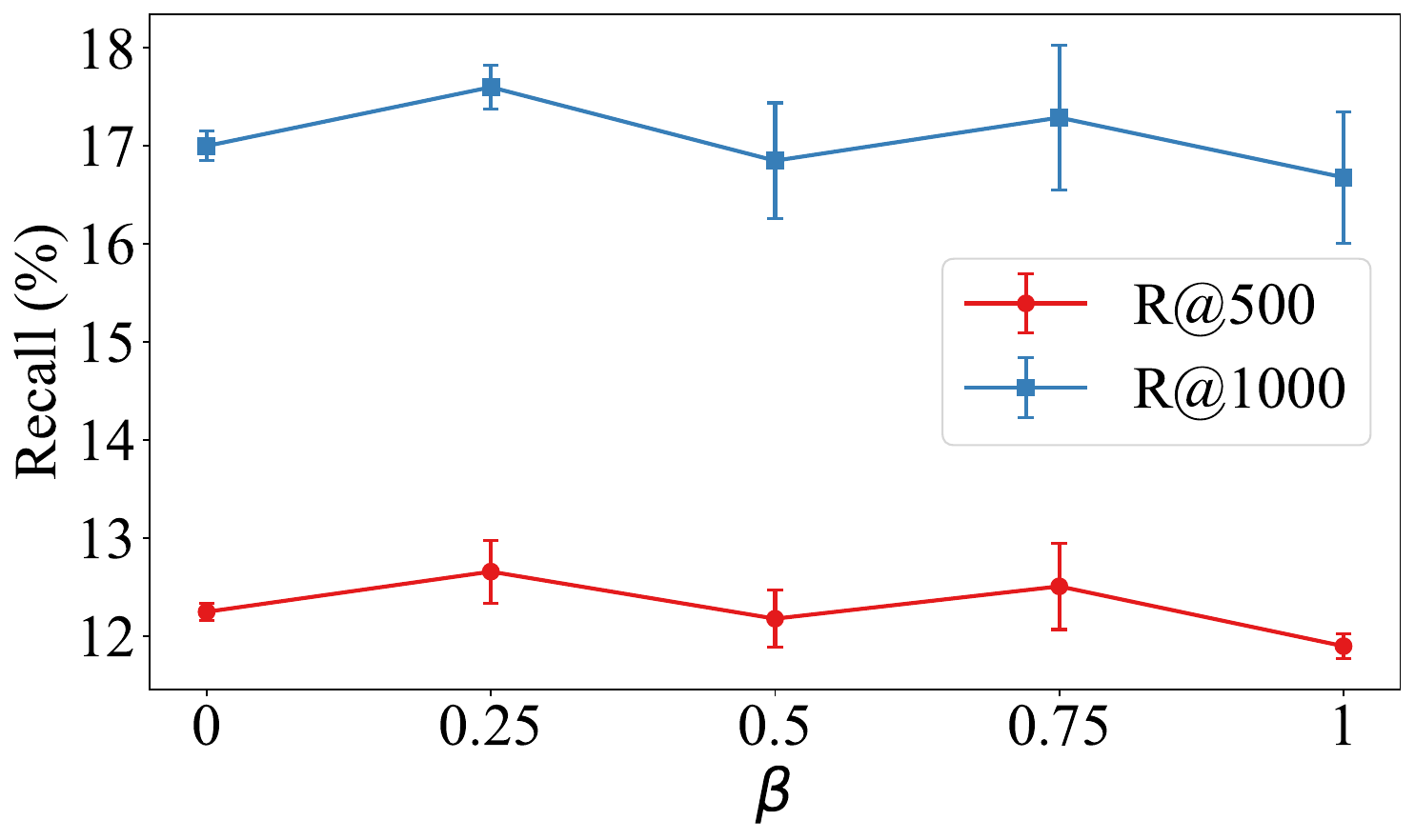}
	}
	\subfigure[Scenario A3]{
		\includegraphics[width=0.225\textwidth]{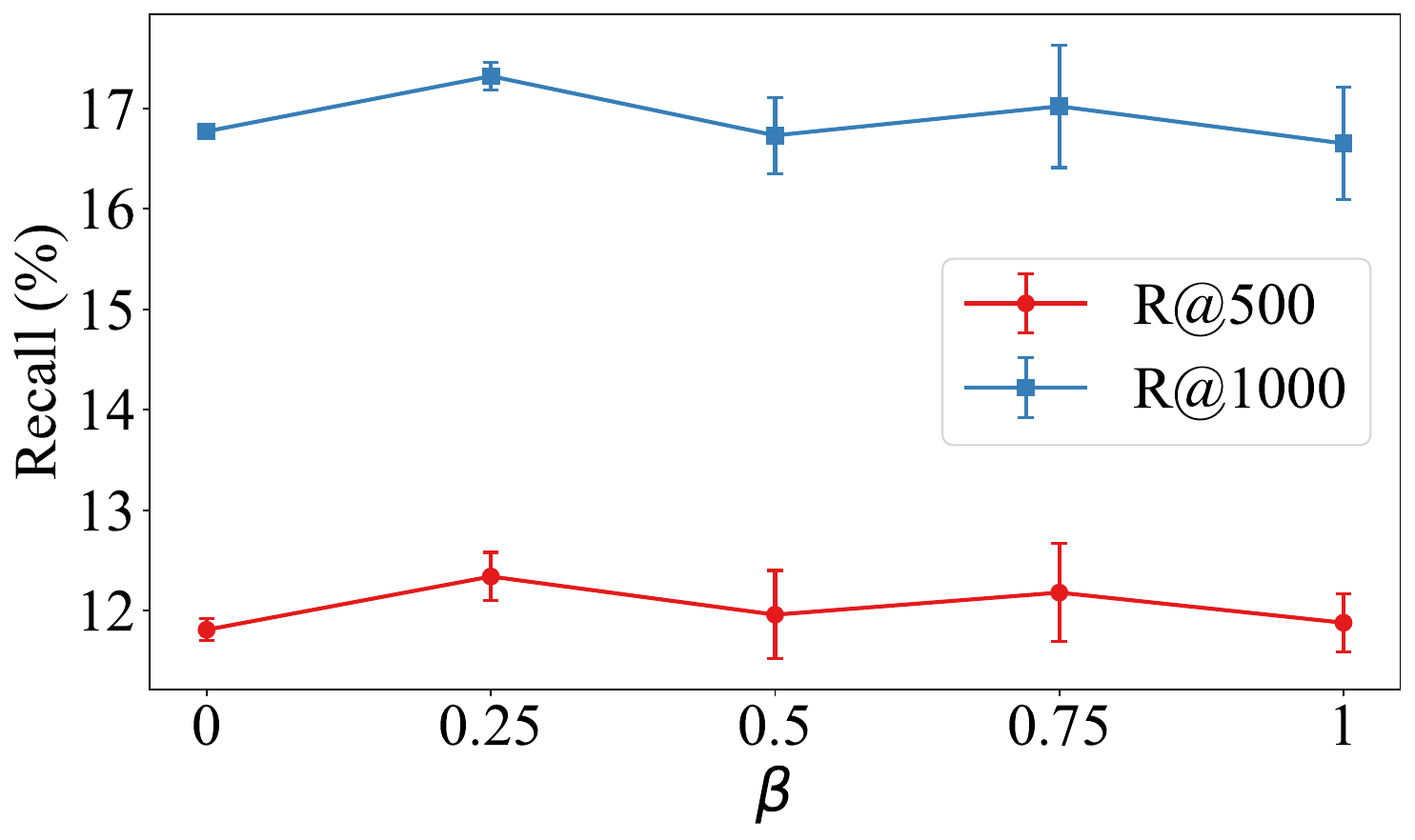}
	}
	\subfigure[Scenario A4]{
		\includegraphics[width=0.225\textwidth]{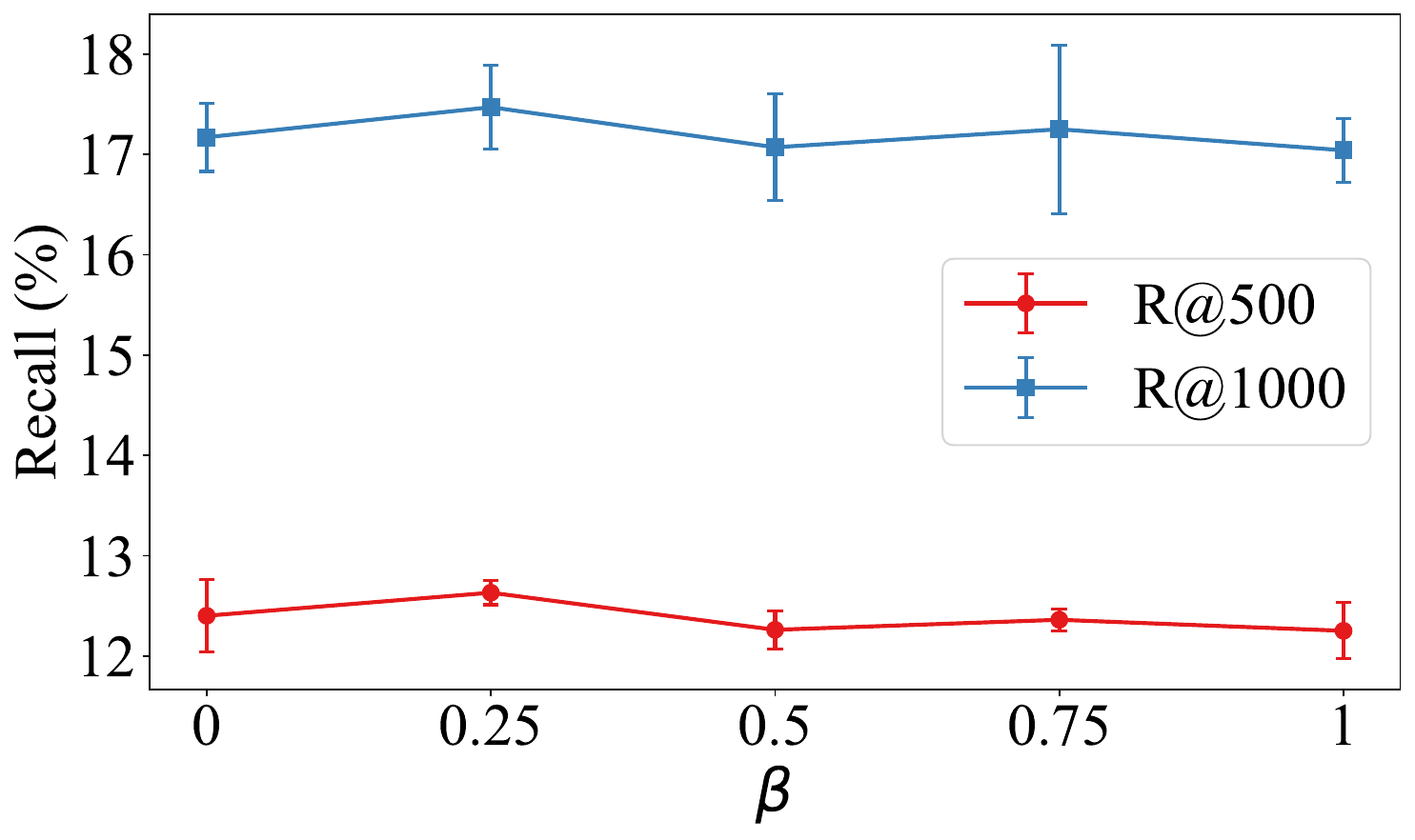}
	}
	\caption{Varying the $\beta$ for PERSCEN in Alimama.}
	\label{fig:betaanalysis}
\end{figure}

\subsubsection{$\beta$ in $\mathcal{L}_\text{VQ}$}
 
$\beta$ in $\mathcal{L}_\text{VQ}$ is an important hyperparameter for the optimization of codebook. 
which is a scalar that trades off the importance of updating hidden vectors and code vectors (e.g. large $\beta$ implies more emphasis on the codebook to adapt towards the encoder)
We explore the impacts of $\beta$, as shown in Figure~\ref{fig:betaanalysis}. 
PERSCEN achieves optimal performance at $\beta=$ 0.25.

\end{document}